\documentclass[prd,aps,floats,twocolumn,showpacs,nofootinbib,preprintnumbers,superscriptaddress]{revtex4}

\usepackage{graphicx}

\usepackage{amsfonts}
\usepackage{amssymb}
\usepackage{amsmath}
\usepackage{mathtools}

\usepackage{color,soul}

\newcommand{\fm}{\,{\rm fm}}

\graphicspath{{.}{Figs/}}                    % Path for graphics

\begin{document}

\preprint{\vbox{\hbox{ADP 14-13/T871, DESY 14-056, LTH 1008, Edinburgh 2014/09}}}

\title{A Feynman-Hellmann approach to the spin structure of hadrons}

\author{A.~J.~Chambers}  \email{alexander.chambers@adelaide.edu.au} \affiliation{CSSM, Department of Physics,
  University of Adelaide, Adelaide SA 5005, Australia}
\author{R.~Horsley}\affiliation{School of Physics and Astronomy,
  University of Edinburgh, Edinburgh EH9 3JZ, UK}
\author{Y.~Nakamura}\affiliation{RIKEN Advanced Institute for
  Computational Science, Kobe, Hyogo 650-0047, Japan}
\author{H.~Perlt} \affiliation{Institut f\"ur Theoretische Physik, Universit\"at Leipzig, 04103 Leipzig, Germany}
\author{D.~Pleiter}\affiliation{JSC, J\"ulich Research Centre, 52425
  J\"ulich, Germany}\affiliation{Institut f\"ur Theoretische Physik,
  Universit\"at Regensburg, 93040 Regensburg, Germany}
\author{P.E.L.~Rakow}\affiliation{Theoretical Physics Division,
  Department of Mathematical Sciences, University of Liverpool,
  Liverpool L69 3BX, UK}
\author{G.~Schierholz}\affiliation{Deutsches Elektronen-Synchrotron
  DESY, 22603 Hamburg, Germany} 
\author{A.~Schiller} \affiliation{Institut f\"ur Theoretische Physik, Universit\"at Leipzig, 04103 Leipzig, Germany}
\author{H.~St\"uben}\affiliation{Regionales Rechenzentrum, Universit\"at Hamburg, 20146 Hamburg, Germany}
\author{R.~D.~Young} 
\author{J.~M.~Zanotti} \affiliation{CSSM, Department of Physics,
  University of Adelaide, Adelaide SA 5005, Australia}
\collaboration{CSSM and QCDSF/UKQCD Collaborations} \noaffiliation


\begin{abstract}

We perform a $N_f=2+1$ lattice QCD simulation to determine the quark
spin fractions of hadrons using the Feynman-Hellmann theorem. By
introducing an external spin operator to the fermion action, the
matrix elements relevant for quark spin fractions are extracted from
the linear response of the hadron energies. 
Simulations indicate that the Feynman-Hellmann method offers statistical precision that is comparable to the 
standard three-point function approach, with the added
benefit that it is less susceptible to excited state
contamination. This suggests that the Feynman-Hellmann technique offers a promising
alternative for calculations of quark line disconnected contributions
to hadronic matrix elements. At the SU(3)-flavour symmetry point,
we find that the connected quark spin fractions are
universally in the range 55-70\% for vector mesons and octet and
decuplet baryons. There is an indication that the amount of spin
suppression is quite sensitive to the strength of SU(3) breaking.

\end{abstract}

\maketitle

\section{Introduction}

The decomposition of the nucleon spin presents a fascinating challenge
for the theoretical understanding of nonperturbative QCD. While the
simplest quark model has all of the nucleon spin attributed to the
spin of its quark constituents, the latest experimental measurements
suggest that only about one third of the nucleon spin comes from the
spin of the quarks \cite{Alexakhin:2006oza}. This has motivated an
extensive theoretical effort to understand the QCD origins of this
quark spin suppression. We refer the reader to the comprehensive
reviews of
Refs.~\cite{Anselmino:1994gn,Filippone:2001ux,Bass:2004xa,Aidala:2012mv}.

Lattice QCD provides a systematically improvable technique to study
nonperturbative features of QCD, and hence offers significant
potential to give valuable insight into the spin decomposition of
the nucleon. Recent results have been published in
\cite{Bratt:2010jn,Syritsyn:2011vk,Sternbeck:2012rw,Alexandrou:2013joa} ---
also see the lattice review \cite{Hagler:2009ni}. Nevertheless, there
are still challenges in the lattice formulation, particularly those
associated with the simulation of so-called ``disconnected''
quantities. Disconnected quantities refer to
those where the external probe couples to a hadron correlator only
through the underlying gauge field configuration.
Standard three-point function techniques require the
stochastic estimation of these quark loop contributions and, while
progress has been made, e.g.\ \cite{QCDSF:2011aa,Deka:2013zha}, it has proven to be
notoriously difficult to extract a non-zero signal.

In the present work we explore an alternative technique for the
extraction of hadronic spin matrix elements in lattice QCD. In
particular, we utilise the Feynman-Hellmann (FH) theorem applied to
the lattice regularisation framework. We consider the energy shifts of
hadrons in the presence of a uniform weak external field which couples
directly to the quark spin. 
This is similar to an idea proposed in \cite{Detmold:2004kw}.
By the FH theorem, the leading linear
response of the energy can be identified with the corresponding spin matrix element
of interest. 
A first exploration of this method was performed in \cite{Horsley:2012}, and later in \cite{Alexandrou:2013tfa}, for the gluon
energy-momentum tensor.
A full simulation would require the generation of new
gauge ensembles which modify the fermion action of the sea quarks, incorporating the
external field. Here we establish the method by coupling the field to
the connected quark fields and benchmark our results against standard
three-point function techniques.

There are some key advantages of the Feynman-Hellmann
method. Importantly, there has been plenty of debate surrounding the
difficulty in controlling excited state contamination in conventional
three-point function calculations of $g_A$ 
\cite{Dinter:2011sg,Capitani:2012gj,Owen:2012ts,Bhattacharya:2013ehc,Jager:2013kha,Bali:2013nla}.
Since the FH method outlined in this paper only requires the
extraction of hadron energies from lattice two-point functions,
greater control of excited state contamination is possible through the
identification of a distinct effective mass plateau.  In addition, the
quark propagators generated in the presence of the external field can
be inserted into any hadronic correlation function and therefore, for
a single set of inversions, one can study the spin content of many
different hadrons. In contrast, usual three-point function methods
require a new sequential propagator for each hadronic state of
interest.\footnote{Nevertheless, the standard three-point method can
  access many different matrix elements simultaneously for a given
  choice of hadronic state.}

With easy access to a variety of hadronic states, we are able to
report first dynamical lattice QCD simulation results for the spin
content of vector mesons and decuplet baryons, in addition to the
baryon octet. Interestingly we find that at the SU(3)-flavour
symmetric point of our simulations the connected quark spin fraction
is around 55-70\%, irrespective of the hadron in question. This is in
line with the general expectation of relativistic corrections to quark
model wave functions
\cite{Thomas:1981vc,Thomas:1982kv,Miller:1984em}. We also present
results away from the SU(3) symmetric point, where we find SU(3)
breaking effects that could lead to significant breakdown of this
universality in the light-quark domain \cite{Shanahan:2013apa}.

The outline of the manuscript is as follows: Section II describes the formalism and
notation used in this paper, and the
strategy for the implementation of the Feynman-Hellmann theorem in
lattice QCD simulations (a detailed derivation of the theorem is
included in Appendix A); the lattice configurations of the present
study are reviewed in Section III; the analysis techniques are
described in Section IV; and our numerical results for various hadrons
are reported in Section V; concluding remarks are summarised in
Section VI.

\section{Formalism}
\label{sec:fhmethod}

In this section we present the formalism and notation used in this paper with 
regard to the spin structure of hadrons, and explain the approach of using the
Feynman-Hellmann theorem to calculate matrix elements.

\subsection{Spin Notation}

We express the total spin of a hadron of spin $J$ by
\begin{equation}
J=\frac12\Delta\Sigma^J+L_q^J+J_G^J,
\end{equation}
where $L_q^J$ and $J_G^J$ denote the 
quark orbital angular momentum and gluon angular momentum,
respectively.
The total quark spin sum is given by
$\Delta\Sigma^J=\sum_q\Delta q^{JJ}$,
which in the Bjorken limit is defined in terms of the zeroth moments of
the polarised quark distributions,
\begin{equation}
\Delta q^{Jm}=\int \mathrm{d} x\left[q^{Jm}_{\uparrow}(x)-q^{Jm}_\downarrow(x)\right].
\end{equation}
Our notation is such that these describe generalisations of polarised
quark distributions for hadrons of spin $J$ with longitudinal spin
polarisation $m$, as defined by Ref.~\cite{Jaffe:1988up}. In lattice
simulations, these can be computed by evaluating matrix elements of
the local operator
\begin{equation}\label{eq:rossEquation}
A_q^\mu=\overline{q}i\gamma_5\gamma^\mu q.
\end{equation}
In the rest frame of the hadron, the forward matrix elements of this operator
directly isolate the quark spin contributions,
\begin{equation}
\langle H,J m|A_q^3(0)|H,Jm\rangle=2M_H\Delta q^{Jm},
\end{equation}
for a hadron with polarisation $m$ with respect to the $z$-axis. It is
these matrix elements that we wish to determine for a variety of
spin-$J$ hadrons, $H$.

\subsection{The Feynman-Hellmann Method}

The Feynman-Hellmann theorem offers an alternative method for
calculating matrix elements of a particular operator. In Appendix
\ref{app:fh} we derive the theorem as relevant for lattice
calculations. Here we summarise the main points.

Suppose we wish to calculate the matrix element of an operator
$\mathcal{O}$ with some hadron state $| H \rangle$.
Consider modifying the QCD action such that
\begin{equation}\label{eq:generalActionModification}
  S \to S + \lambda \int \mathrm{d}^4 x \, \mathcal{O}(x) \, ,
\end{equation}
where $\lambda$ is a real parameter, and $\mathcal{O}$ is a local operator.
Then by the Feynman-Hellmann theorem we have that
\begin{equation}
  \frac{\partial E(\lambda)} {\partial \lambda} = \frac{1}{2
    E(\lambda)} \langle H | \mathcal{O} | H \rangle_\lambda \, ,
\end{equation}
where $E$ is the energy of the hadron state, and the subscript $\lambda$ on
the correlator indicates that it is evaluated with respect to the
modified action. Note when $\lambda = 0$, we have
\begin{equation}\label{eq:important}
  \left. \frac{\partial E(\lambda)} {\partial \lambda}
  \right|_{\lambda=0} = \frac{1}{2 E} \langle H | \mathcal{O} | H
  \rangle \, ,
\end{equation}
where the matrix element on the right-hand side is now with respect to
the unmodified action.
If the modification in
Eq.~(\ref{eq:generalActionModification}) is made in the context of a lattice
calculation, then one can examine the behaviour of hadron energies as
the parameter $\lambda$ changes, and extract the above matrix element
at the point where $\lambda=0$.

Recall the lattice estimate of the expectation value of an operator $\mathcal{O}$ over field configurations $U^{(i)}$ is given by
\begin{equation}
  \langle \mathcal{O} \rangle \approx \frac{1}{N} \sum_{i=1}^N
  \overbracket[0.5pt]{\mathcal{O}}[U^{(i)}] \, ,
\end{equation}
where the bracket over $\mathcal{O}$ indicates that all quark bilinears in $\mathcal{O}$ have been
 Wick contracted and replaced with quark propagators, and where the field configurations have been
generated using the weighting
\begin{equation}\label{eq:weighting}
  \det[D(U)]e^{-S_g[U]}\, .
\end{equation}
There are two points at which modifications to the action may be made in this calculation.

Firstly, quark propagators in the operator $\mathcal{O}$ are calculated by inverting the Dirac
operator matrix.
This matrix is given by the quark contribution to the
QCD action, and so must be modified if we change the quark action.
This change is straightforward to apply, only requiring a redefinition of the
Dirac operator.

Secondly, we note that the weighting of the gauge fields in Eq.~(\ref{eq:weighting}) 
depends on both the quark component of the
action in the functional determinant, and the gluon component in the
exponential.
Hence, any modification we make to the action should be included
during the generation of the background gauge fields.

By choosing to neglect either one of these modifications, we are able
to individually isolate connected and disconnected contributions to matrix elements.
Modifications to the gauge configurations allow access to disconnected quantities,
and modifications to the calculation of propagators allow access to connected quantities.

The method above presents several advantages for calculating, in particular, quantities such as the
disconnected quark contributions to the proton spin.
Such disconnected contributions
are included in a simulation during the generation of gauge configurations,
and the calculation of the appropriate matrix element
is reduced to the calculation of hadron energies for different
values of $\lambda$, in order to apply Eq.~(\ref{eq:important}).

We will demonstrate the implementation of the Feynman-Hellmann method by 
calculating  the connected quark contributions to the spin of hadrons. This has been investigated 
previously using standard three-point function methods, results with which we will
compare our calculations.

The simulations discussed in Sec.~\ref{sec:connectedSpin} and
Sec.~\ref{sec:results} make use of the partially quenched case
for calculating connected quantities, and we do not generate any
modified field configurations in the present work.

\section{Simulation Details}
\label{sec:latticeSetup}

We use gauge field configurations with $2+1$ flavours of
non-perturbatively $\mathcal{O}(a)$-improved Wilson fermions and a
lattice volume of $L^3 \times T = 32^3 \times 64$. The lattice spacing
$a = 0.074(2) \fm$ is set using a number of singlet quantities
\cite{Horsley:2013,Bietenholz:2010,Bietenholz:2011}. The clover action
used comprises the tree-level Symanzik improved gluon action together
with a stout smeared fermion action, modified (as described in
Sec.~\ref{sec:connectedSpin}) for the implementation of the
Feynman-Hellmann method. We have ensembles with two sets of hopping
parameters, $(\kappa_l,\kappa_s) = (0.120900,120900),(0.121040,120620)$, where we work
in the isospin-symmetric limit such that $\kappa_l = \kappa_u = \kappa_d$.
 Table~\ref{tab:ensembleDetails} gives the masses of various
hadrons as realised on these configurations \cite{Bietenholz:2011}.

\begin{table}
  \begin{tabular}{c |c c}
    \hline \hline $\kappa_l$ & 0.120900 & 0.121040 \\ $\kappa_s$ &
    0.120900 & 0.120620 \\ \hline $aM_\pi$ & 0.1747(5) & 0.1349(5)
    \\ $aM_N$ & 0.4673(27) & 0.4267(50) \\ $aM_\Lambda$ & 0.4673(27) &
    0.4547(43) \\ $aM_\Delta$ & 0.5676(64) & 0.5520(79) \\ $aM_\rho$ &
    0.3341(34) & 0.3127(38) \\ \hline \hline
  \end{tabular}
  \caption{Table of hadron masses (in lattice units) for each
    ensemble.}
  \label{tab:ensembleDetails}
\end{table}

As discussed in the next section, the initial investigation of this
method is performed at the SU(3) symmetric point ($\kappa_l = \kappa_s
= 0.120900$) where all three quarks have the same mass, corresponding to a pion mass of
 around 470 MeV. On a subset of 350 configurations we
explore the feasibility of the method using up to four different
values of $\lambda$.

After tuning the method at this point, we then apply it to an ensemble
with a lighter pion mass of around 360 MeV. As all of our lattice
ensembles are generated with the singlet quark mass $\overline{m} =
\frac{1}{3}(2 m_l + m_s)$ held fixed, this lattice also contains
a heavier strange quark. This will allow us to demonstrate the
suitability of this method for the study of the quark spin
contributions to a variety of hadrons.

Unless otherwise stated, all results quoted in the remainder of this
paper are unrenormalised (indicated by a superscript `latt.'). 
However in order to compare with existing results in the literature, 
we use preliminary results for the non-singlet axial current renormalisation 
constant \cite{axialRenormalisation}
\begin{equation}\label{eq:axialRenormalisation}
Z^{\mathrm{NS}}_A=0.85(2) \, .
\end{equation}
We note that most of the results quoted in the remainder of the paper are either for the total or individual quark spin contributions to a hadron's spin which also requires knowledge of the singlet axial current renormalisation $Z_A^{\mathrm{s}}$ \cite{QCDSF:2011aa}, which has an anomalous dimension. Since $Z_A^{\mathrm{S}}$ deviates from $Z^{\mathrm{NS}}_A$ starting at $\mathcal{O}(\alpha_s^2)$ in perturbation theory, we expect $Z_A^{\mathrm{S}}$ in the $\overline{\mathrm{MS}}$ scheme at $\mu^2=4\,\mathrm{GeV}^2$ to differ from $Z_A^{\mathrm{NS}}$ by no more than a couple of percent. A similar sized correction maybe be needed to achieve full $\mathcal{O}(a)$ improvement \cite{Capitani:2000xi}. In future work where we intend to also include disconnected contributions, we will implement a proper treatment of the renormalisation. However, for the exploratory work carried out in this paper, we neglect these minor corrections and simply use $Z^{\mathrm{NS}}_A$  in Eq.~(\ref{eq:axialRenormalisation}) when a comparison of renormalised results is made.

\section{Analysis Techniques}
\label{sec:connectedSpin}

Here we show how the Feynman-Hellmann theorem may be applied to
calculate quark axial charges of hadrons, using the proton as an
example.  We will then show how the determination of these axial
charges can be improved through the use of ratios of lattice two-point
functions.  Finally we will investigate the optimal
choice of $\lambda$ values needed to reliably determine the axial
charges at minimal computational cost.

\subsection{Spin Operator \& Spin Projection}

In our simulations, we modify the QCD action such that
\begin{equation}\label{eq:actionModification}
  S \to S(\lambda) = S + \lambda \sum_x \overline{q}(x) i\gamma_5\gamma_3
  q(x) \, ,
\end{equation}
where $q$ denotes a particular quark flavour. Note $i \gamma_5
\gamma_3$ is the Euclidean-space form of the spin operator in the
$z$-direction.
By application of the Feynman-Hellmann theorem for a zero-momentum
hadron $H$ we have
\begin{equation}\label{eq:applicationOfFeynmanHellmann}
  \left. \frac{\partial E(\lambda)} {\partial \lambda}
  \right|_{\lambda=0} = \frac{1}{2 M} \langle H | \overline{q}
  i\gamma_5\gamma_3 q | H \rangle \, .
\end{equation}
Comparing with Eq.~(\ref{eq:rossEquation}), we see that this slope gives direct access to the quark spin contributions,
\begin{equation}\label{eq:thingy}
   \Delta q = \left. \frac{\partial E(\lambda)} {\partial \lambda}
   \right|_{\lambda=0} \, .
\end{equation}
For simplicity, we have suppressed the explicit $J$ and $m$ spin indices, as is conventional
for a spin-$\frac{1}{2}$ target.
Calculation of $\Delta q$ has now been reduced from the calculation of
lattice three-point functions to the simpler task of measuring
energies from lattice two-point functions.

In our simulations, the modification to the action in
Eq.~(\ref{eq:actionModification}) is only made to the Dirac matrix
when calculating propagators, hence we only access the quark connected
contributions to $\Delta q$, as discussed in Sec.~\ref{sec:fhmethod}.
Hence on the lattice, we have that
\begin{equation}\label{eq:onTheLattice}
  {\Delta q}_\text{conn.}^\text{latt.} = \left. \frac{\partial
    E(\lambda)} {\partial \lambda} \right|_{\lambda=0} \, .
\end{equation}

Calculation of proton energies proceeds via normal lattice
hadron-spectroscopy techniques.
We make use of the standard proton interpolating operator
\begin{equation}\label{eq:protonInterpolator}
  \mathcal{O}_p = \epsilon_{abc} \left( u_a^T C \gamma_5 d_b \right)
  u_c \, ,
\end{equation}
where only colour indices are shown explicitly; spinor indices are
implied by matrix/vector notation.
We use the positive parity projection operator (in Euclidean space)
\begin{equation}\label{eq:parityProjectionOperators}
  \Gamma_4 = \frac{1}{2} \left( 1 + \gamma_4 \right)
\end{equation}
to project out the positive parity state. Since the matrix element in
Eq.~(\ref{eq:rossEquation}) requires the hadron state to have definite spin, we combine the operator in Eq.~(\ref{eq:parityProjectionOperators}) with spin-projection operators, 
\begin{equation}\label{eq:spinProjectionOperators}
  \Gamma_\pm = \frac{1}{2} (1 \pm i \gamma_5 \gamma_3) \Gamma_4.
\end{equation}
Together these operators allow us to project out the $m = \pm \frac{1}{2}$ positive-parity proton states.

Recalling Eq.~(\ref{eq:actionModification}), we note that reversing the spin
polarisation of the hadron state is equivalent to reversing the sign
of $\lambda$.
Hence with a single choice of $\lambda$ we are able to identify the
energies of the spin-up proton with positive $\lambda$, and those of
the spin-down proton with negative $\lambda$.
In this way, we effectively double our sampled parameters, without
increasing the simulation time.

\begin{figure}
  \includegraphics[width=\columnwidth]{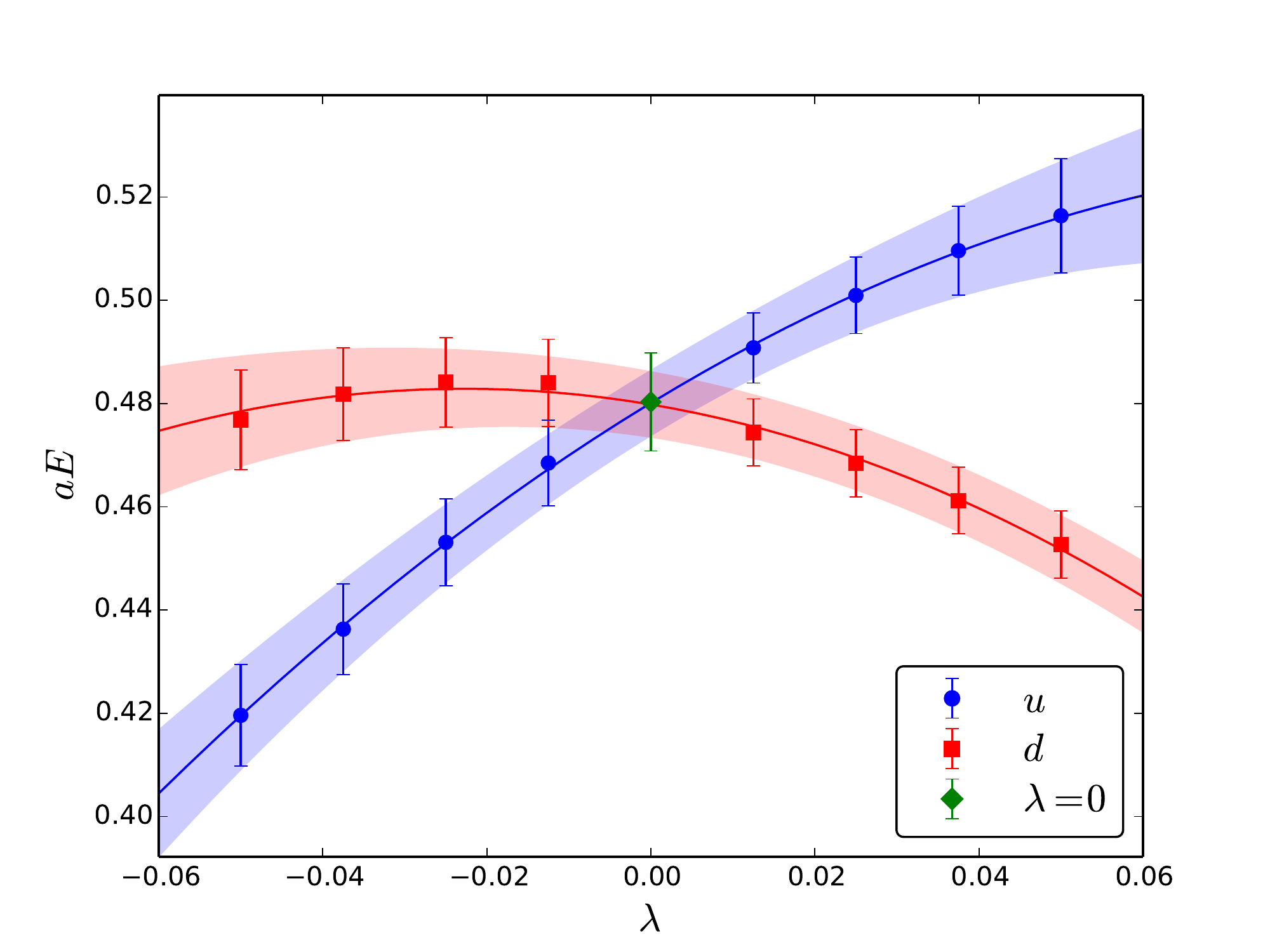}
  \caption{Change in proton energy with the parameter $\lambda$. The
    two datasets show the effect when the extra term is applied to
    each light flavour. Note that at the $\lambda=0$ point we have the
    unshifted proton energy averaged over both spin projections. $\kappa_l = \kappa_s = 0.120900$. }
  \label{fig:nucleonLambdaMass}
\end{figure}
As a first test, we simulate with four values of $\lambda = 0.0125,
0.025, 0.0375, 0.05$ at the SU(3)-flavour symmetric point ($\kappa_l =
\kappa_s = 0.120900$).  Fig.~\ref{fig:nucleonLambdaMass} shows
results for the ground state proton energy as a function of $\lambda$ for
both spin-up (positive $\lambda$) and down (negative $\lambda$)
states.  In the two datasets, the $\lambda$ term in
Eq.~(\ref{eq:actionModification}) has been added to the up quark and
down quark separately.  We fit to a Taylor expansion in the parameter
$\lambda$,
\begin{equation}\label{eq:lambdaFit}
  E(\lambda) = E(0) + \lambda \Delta q + \left. \frac{1}{2} \lambda^2
  \frac{\partial^2 E(\lambda)}{\partial \lambda^2}
  \right|_{\lambda=0} + \ldots \, ,
\end{equation}
retaining only up to quadratic terms in this case.
We see that the slopes of the energy as a function of $\lambda$ for the two flavours
of quark have opposite
signs, indicating the expected result that the up quark has a positive
spin contribution, and the down quark a negative contribution.
We also note the presence of quadratic and higher order terms in
$\lambda$ at larger $\lambda$. These are not presently of interest, as
only the linear behaviour at $\lambda=0$ is required to apply
Eq.~(\ref{eq:onTheLattice}).

Using the linear parameter from the fit in Eq.~(\ref{eq:lambdaFit}),
we have for the (unrenormalised) connected quark spin contributions in
the proton,
\begin{align}
  {\Delta u}_\text{conn.}^\text{latt.} & = 0.97(13) \, , \label{eq:crappyResult1} \\ 
  {\Delta d}_\text{conn.}^\text{latt.} & = -0.27(11) \, . \label{eq:crappyResult2}
\end{align}
The errors here come from a bootstrap analysis of the proton
correlators. Noting the form of the interpolating operator in
Eq.~(\ref{eq:protonInterpolator}), specifically the quark content, we
may interchange up and down quarks above to obtain symmetric results
for the proton's isospin partner, the neutron. Henceforth we will not
distinguish between individual members of isospin multiplets when
quoting results (all calculations are performed in the
isospin symmetric limit). Specific quark flavours can be deduced from
the context.

\subsection{Correlator Ratios}\label{subsec:ratios}
By taking advantage of the correlation between results at different
$\lambda$ using the same statistical ensemble, we may dramatically
improve the previous results.
We can write the energy of a general hadron in terms of an energy shift $\Delta E (\lambda)$ as
\begin{equation}
  E(\lambda) = E(\lambda = 0) + \Delta E(\lambda) \, ,
\end{equation}
where $E_0 = M$ is the mass of the hadron.
Then Eq.~(\ref{eq:onTheLattice}) becomes
\begin{equation}
  {\Delta q}_\text{conn.}^\text{latt.} = \left. \frac{\partial \Delta
    E(\lambda)} {\partial \lambda} \right|_{\lambda=0} \, .
\end{equation}
Hence we only need to calculate energy shifts with respect to
$\lambda$ in order to make use of the Feynman-Hellmann theorem.  These
energy shifts can be determined accurately from ratios of two-point
functions.

For large times $t$ we expect that a lattice two-point function has the
asymptotic form
\begin{equation}
  C(\lambda,t) \overset{\text{large } t}{\longrightarrow} \frac{e^{-E(\lambda) t}}{2
    E(\lambda)} | A(\lambda) |^2 \, .
\end{equation}
Considering the ratio of two such correlation functions, one
calculated with $\lambda=0$ and the other at $\lambda \ne 0$, we have
\begin{equation}\label{eq:finalRatioExpression}
  \frac{C(\lambda,t)}{C(\lambda=0,t)} \overset{\text{large } t}{\longrightarrow}
  e^{-\Delta E(\lambda) t} \frac{E(0)}{E(\lambda)} \frac{| A(\lambda)
    |^2}{| A(0) |^2}.
\end{equation}
The exponential dependence of the above ratio of correlators
contains the difference in energies between the undisturbed energy and
the energy at some $\lambda$.
Using this quantity to measure energy shifts allows us to make use of
correlations between calculations with different values of $\lambda$.
Since each calculation is performed using the same set of underlying
gauge configurations, we expect fluctuations in the correlators to
largely cancel, leaving a much cleaner signal.

\begin{figure}
  \includegraphics[width=\columnwidth]{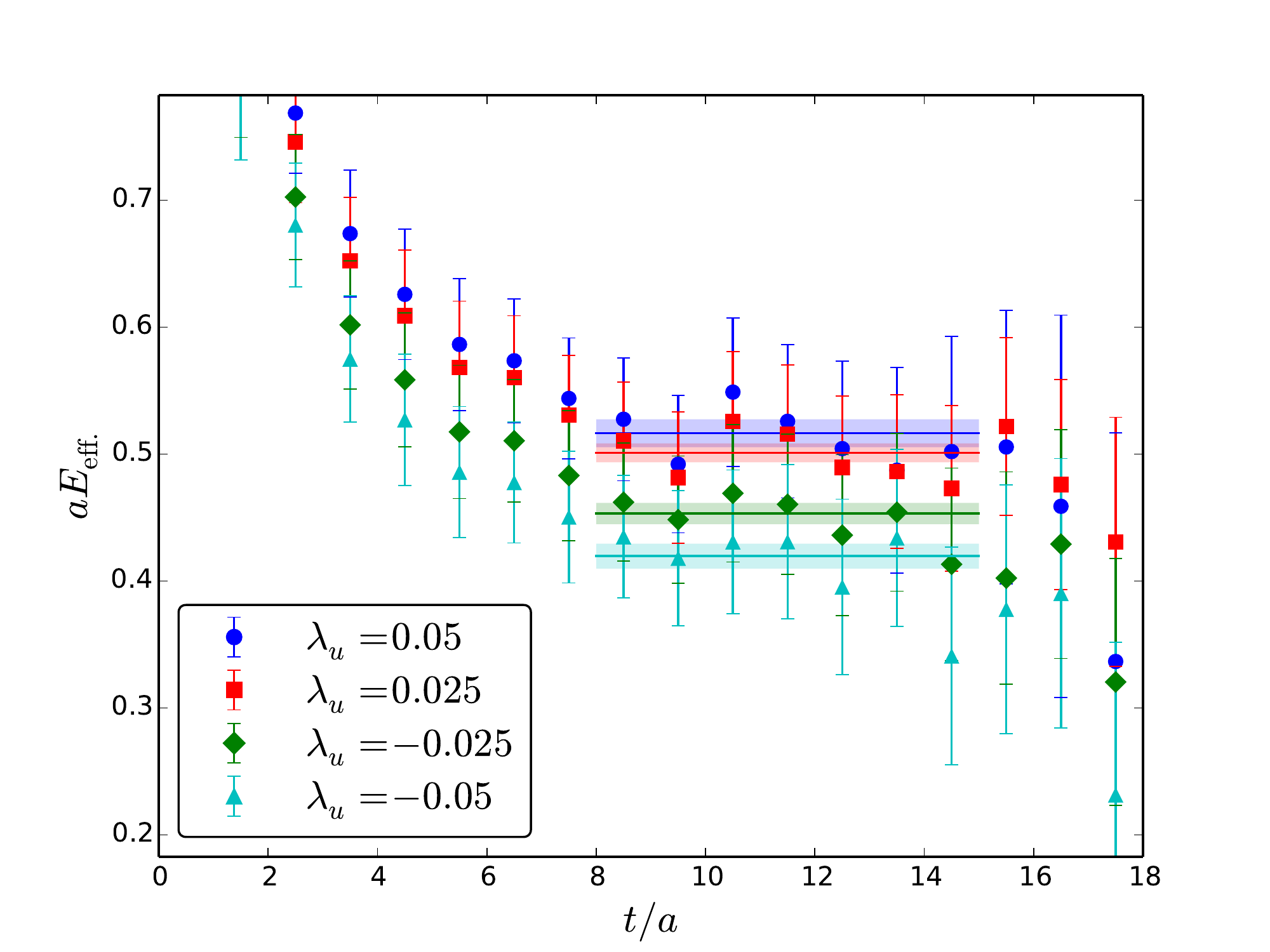}
  \includegraphics[width=\columnwidth]{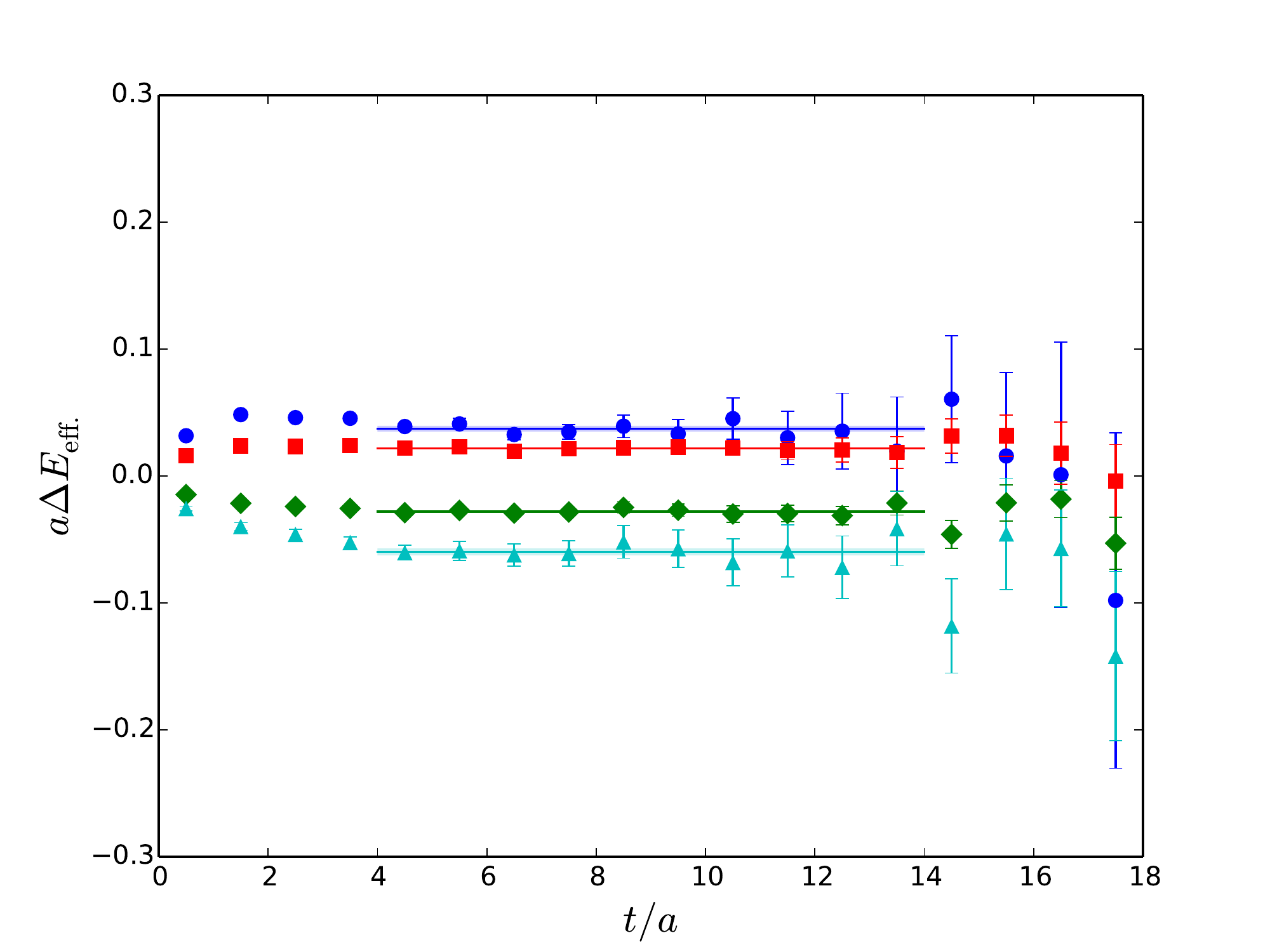}
  \caption{Nucleon effective mass plots for different values of
    $\lambda$ on the up quark at the SU(3)-flavour symmetric
    point. The first plot shows absolute energies, and the second
    energy shifts obtained from correlator ratios. Only a few
    $\lambda$ have been included for clarity. The vertical scale is
    the same for both plots (only shifted), emphasising the
    improvement achieved.}
  \label{fig:nucleonMassSplitting}
\end{figure}
Returning to the example of the last section, the upper plot in
Fig.~\ref{fig:nucleonMassSplitting} shows nucleon effective mass plots
for different values of $\lambda$ on the up quark, and the fit-range
used for each. The lower plot displays the effective masses for the energy shifts
obtained from the ratio of correlators in Eq.~(\ref{eq:finalRatioExpression}). We note that the energy shifts are much clearer using the
new procedure, and we are able to fit at earlier times, possibly due
to the cancelling of excited states.
\begin{figure}
  \includegraphics[width=\columnwidth]{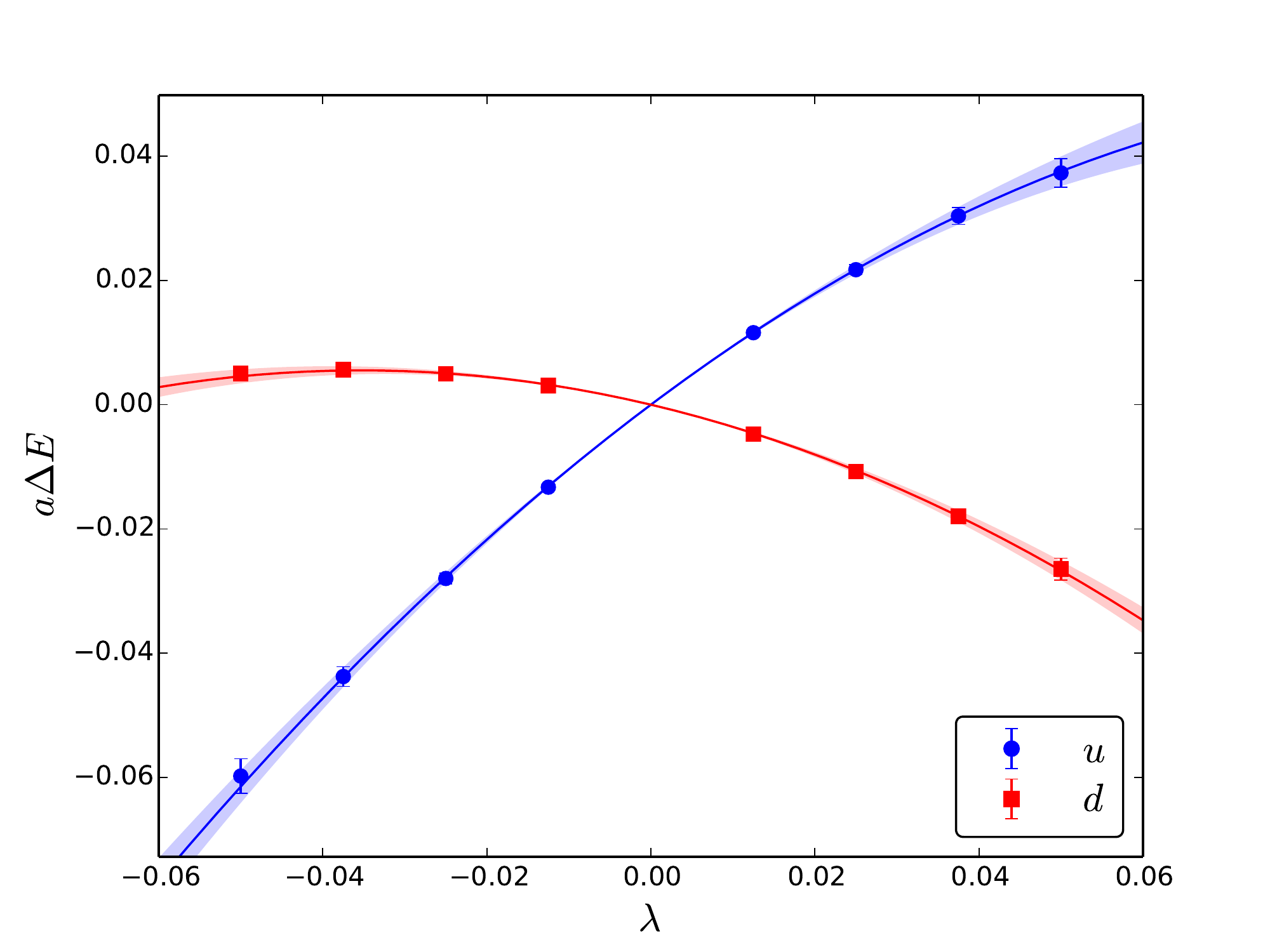}
  \caption{Change in nucleon energy for different parameter values
    with a quadratic fit. $\kappa_l = \kappa_s = 0.120900$.}
  \label{fig:nucleonLambdaRatio}
\end{figure}
Fig.~\ref{fig:nucleonLambdaRatio} shows the resulting nucleon energy shifts as a
function of $\lambda$. We observe that the relative errors between different points
are now much smaller, and we are able to much more tightly constrain the
quadratic fit as compared to Fig.~\ref{fig:nucleonLambdaMass}. We are
also able to fix the $\lambda = 0$ point to zero, since there is no
energy shift for zero background spin-field. The linear plus quadratic
behaviour seen previously has been preserved, as we have only shifted
all data points by a constant amount.

From the linear parameter in the quadratic fit, we calculate the quark
axial charges
\begin{align}
  {\Delta u}_\text{conn.}^\text{latt.} & = 0.990(20) \, , \label{eq:betterResults1} \\
  {\Delta d}_\text{conn.}^\text{latt.} & = -0.313(14) \, .  \label{eq:betterResults2}
\end{align}
These results are consistent with
those in Eq.~(\ref{eq:crappyResult1}) and Eq.~(\ref{eq:crappyResult2}), and our uncertainties have been
significantly reduced.
Note that if we instead extract the linear parameter from a fit including also a cubic term in Eq.~(\ref{eq:lambdaFit}), we find
no change within the quoted statistical error, suggesting that retaining terms up to quadratic order in $\lambda$ is sufficient.

For comparison we have results calculated with a three-point function
method \cite{Cooke:2013} using 330 configurations from the same
larger ensemble of 1500 configurations from which the 350 used in this
work were sourced. The three-point functions method gives for the quark axial charges,
\begin{align}
  {\Delta u}_\text{conn.}^\text{latt.} & = 0.911(29) \, ,\\ {\Delta
    d}_\text{conn.}^\text{latt.} & = -0.290(16) \, ,
\end{align}
where we see comparable precision with our results in Eq.~(\ref{eq:betterResults1}) and Eq.~(\ref{eq:betterResults2}), but obvious tension with 
the result for the up quark. This may be due to the fact that the Feynman-Hellmann method has a greater control of excited state contamination than the 
fixed-sink three-point method with a single source-sink smearing.

\subsection{Optimisation}\label{subsec:optimisation}
The spin matrix elements studied here have utilised numerous values of
the background field strength $\lambda$ in order to accurately
determine the derivative in the zero-field limit. As each value of the
background field parameter requires the computation of a new set of
propagators, we explore how one could best optimise the signal
strength for a minimal set of inversions. This optimisation is
particularly necessary in the context of extending this work to
disconnected operators, where new additional simulations are required for each value of the field strength.

We would like to realise a minimum of two different field strengths
(with spin-up/down projections), and restrict ourselves
to a fixed-intercept quadratic fit in $\lambda$.  Quadratic terms do not
affect the linear terms that we are interested in, because these terms
shift the energies equally on either side of the $\lambda=0$
point. Realising a minimum of two field strengths
(four values of $\lambda$, after spin-up/down projection) allows us to be confident in 
uncertainties calculated for the two-parameter fixed-intercept quadratic fit.

We consider fitting quadratically to subsets of our existing results at the SU(3)-flavour symmetric point ($\kappa_l = \kappa_s = 0.120900$), realising only
two of the four values of $\lambda$ at a time (four total data points after spin projection).
Table~\ref{tab:optimisation} shows results for the quark axial charges
calculated using these subsets.
\begin{table}
  \begin{tabular}{c c | c c }
    \hline \hline $\lambda_1$ & $\lambda_2$ & ${\Delta u}_\text{conn.}^\text{latt.} $ & ${\Delta d}_\text{conn.}^\text{latt.}$ \\ 
    \hline 
    0.0125 & 0.0250 & 0.994(18) & -0.313(13) \\ 
    0.0125 & 0.0375 & 0.992(19) & -0.312(13) \\ 
    0.0125 & 0.0500 & 0.988(19) & -0.311(14) \\ 
    0.0250 & 0.0375 & 0.991(21) & -0.314(14) \\
    0.0250 & 0.0500 & 0.987(23) & -0.313(15) \\
    0.0375 & 0.0500 & 0.981(27) & -0.314(17) \\
    \hline
    \hline
  \end{tabular}
  \caption{Connected spin contributions to the proton calculated using partial fits
    to only two values of the external field strength, $\lambda$. The fit
    used is given in Eq.~(\ref{eq:lambdaFit}), where we retain up to quadratic terms.
    The first column lists the values of $\lambda$ used, and the second
    shows the calculated values of the quark axial charges that
    result.}
  \label{tab:optimisation}
\end{table}
The calculated axial charges remain consistent within errors with each
different fit, however there does seem to be a systematic shift in $\Delta u$ as we move to
higher $\lambda$.  Also, the statistical errors in the energy
shifts increase as $\lambda$ increases.  Importantly, we note that
the results obtained from a quadratic fit to the smallest two values
of $\lambda$ (in the top row of Table~\ref{tab:optimisation}) agree within errors with results obtained from a fit to the entire data set in Eq.~(\ref{eq:betterResults1}) and
Eq.~(\ref{eq:betterResults2}), and with comparable statistical
error. For these reasons, we restrict ourselves
to realising only the two smallest $\lambda$ when considering our second ensemble at smaller pion mass, as this appears to give
sufficient accuracy and precision.

Also, we note that using only two values of the background field strength
brings the total number of matrix inversions required for a computation of the axial charge of the proton in line
with the standard three-point function method.
With two values of $\lambda$, three colours and four spinor indices, we must calculate 36 inversions 
for every 
operator that we wish to investigate.
For a three-point function calculation, three colours, four spinor indices and three quark propagators also
lead to 36 inversions for each hadron that we consider.
If the aim is to compute the forward proton matrix element of the axial operator, our results indicate
that the Feynman-Hellmann method can achieve comparable statistical precision to the three-point function approach,
at fixed computational cost.

\section{Results}\label{sec:results}

Here we summarise connected quark spin contributions obtained using
correlator ratio methods for the octet and decuplet baryons, and
vector mesons. 
All results quoted are from quadratic fits in $\lambda$.
Calculations at the SU(3) symmetric point make use of the full dataset of four values of $\lambda$. Simulations carried out away from the SU(3)-flavour symmetric point realise two values of the background field strength parameter, $\lambda = 0.0125, 0.025$, as motivated by the discussion in Sec.~\ref{subsec:optimisation}. For all analyses we make use of correlator ratios as discussed in Sec.~\ref{subsec:ratios}.

\subsection{Octet Baryons $(J = \frac{1}{2})$}

Using the preliminary renormalisation in
Eq.~(\ref{eq:axialRenormalisation}), we conclude our discussion of the
proton in Sec.~\ref{sec:connectedSpin} by quoting renormalised values for $g_A$ at our 
two simulated pion masses,
\begin{align}
  g_A (m_\pi = 470 \text{MeV}) & = 1.105(29) \, , \\
  g_A (m_\pi = 360 \text{MeV}) & = 1.072(32) \, ,
\end{align}
which are in good agreement with results in the literature, \cite{Dinter:2011sg,Capitani:2012gj,Owen:2012ts,Bhattacharya:2013ehc,Jager:2013kha,Bali:2013nla} 
(or \cite{Syritsyn:2014saa} for a recent review).
For the remaining octet baryons (excluding the $\Lambda$) we re-use
the form of the interpolating operator for the proton in
Eq.~(\ref{eq:protonInterpolator}),
\begin{equation}\label{eq:octetInterpolator}
  \mathcal{O}_\text{octet} = \epsilon_{abc} \left( {q_1}_a^T C
  \gamma_5 {q_2}_b \right) {q_1}_c \, ,
\end{equation}
substituting light and strange quarks to access the $\Sigma$ and $\Xi$
states (in the isospin-symmetric limit). For example for the
$\Sigma^+$ we use the operator
\begin{equation}
  \mathcal{O}_{\Sigma^+} = \epsilon_{abc} \left( {u}_a^T C \gamma_5
          {s}_b \right) {u}_c \, .
\end{equation}
In addition, we use the spin and parity-projection operators given in
Eq.~(\ref{eq:parityProjectionOperators}) and
Eq.~(\ref{eq:spinProjectionOperators}).

The calculation proceeds as described in Sec.~\ref{sec:connectedSpin}, and
Table~\ref{tab:octetResults} shows results for the octet (details of
the $\Lambda$ calculation are discussed later).
\begin{table}
  \begin{tabular}{c|c c c c}
    \hline 
    \hline 
    $\kappa_l$ & \multicolumn{2}{c}{0.120900} & \multicolumn{2}{c}{0.121040} \\
    $\kappa_s$ & \multicolumn{2}{c}{0.120900} & \multicolumn{2}{c}{0.120620} \\
    \hline 
    & ${\Delta q_1}_\text{conn.}^\text{latt.}$ & ${\Delta q_2}_\text{conn.}^\text{latt.}$ & ${\Delta q_1}_\text{conn.}^\text{latt.}$ & ${\Delta q_2}_\text{conn.}^\text{latt.}$ \\
    \hline 
    N & 0.990(20) & -0.313(14) & 0.971(22) & -0.291(20) \\
    $\Sigma$ & 0.990(20) & -0.313(14) & 0.948(18) & -0.297(8) \\
    $\Xi$ & 0.990(20) & -0.313(14) & 1.039(12) & -0.275(11) \\
    $\Lambda$ & -0.070(23) & 0.785(18) & -0.050(17) & 0.803(10) \\
    \hline
    \hline
  \end{tabular}
  \caption{Table of connected spin contributions for the baryon
    octet. For all baryons except the $\Lambda$, $q_1$ and $q_2$ are
    as they appear in Eq.~(\ref{eq:octetInterpolator}). For the
    $\Lambda$, $\Delta q_1 = \Delta u + \Delta d$ and $\Delta q_2 =
    \Delta s$.}
  \label{tab:octetResults}
\end{table}
$q_1$ and $q_2$ in the table refer to the quark flavours as they
appear in the appropriate form of the interpolating operator in
Eq.~(\ref{eq:octetInterpolator}). As we are working in the
isospin-symmetric limit, the results quoted can be applied to all
members of each isospin multiplet, with appropriate flavour
re-labelling. So for instance we have for the $\Sigma^+$ (quark
content $uus$) that $\Delta q_1 = \Delta u$ and $\Delta q_2 = \Delta
s$, whereas for the $\Xi^0$ (quark content $uss$) $\Delta q_1 = \Delta
s$, $\Delta q_2 = \Delta u$.

Away from the SU(3) symmetric point (at the lighter pion mass) we see
evidence for SU(3)-flavour-breaking effects in the quark spin contributions to the baryon octet. As we
discussed in Sec.~\ref{sec:latticeSetup}, the singlet quark mass is
the same for both ensembles, so the light quarks are lighter and the
strange quark heavier on the second ensemble. We see $\Delta u$ and
$\Delta d$ decreasing for the nucleon, whereas $\Delta u$ $(\Delta d)$
decreases and $\Delta s$ increases for the $\Xi^{0(-)}$.
 
By comparing the individual quark flavour results of the octet baryons, we can gain an insight into the environmental sensitivity of the
quark axial charges. As we move from the N to $\Sigma$ state with the
heavier strange quark for example, we see the light quark contribution
decreasing.

For the $\Lambda$ baryon we use the interpolating operator
\begin{align}\label{eq:lambdaInterpolator}
  \mathcal{O}_{\Lambda} = \epsilon_{abc} \frac{1}{\sqrt{6}} \big[ 2
    \left( {u}_a^T C \gamma_5 {d}_b \right) {s}_c & + \left( {u}_a^T C
    \gamma_5 {s}_b \right) {d}_c \\ \notag & - \left( {d}_a^T C
    \gamma_5 {s}_b \right) {u}_c \big] \, .
\end{align}
Note that when calculating two-point functions for the $\Lambda$, we
do not calculate separate propagators for the up and down
quarks. Hence the spin-field term in Eq.~(\ref{eq:actionModification})
is added to both light-quarks at once, and so in
Table~\ref{tab:octetResults}, $\Delta q_1 = \Delta u + \Delta d$ and
$\Delta q_2 = \Delta s$.
\begin{figure}
  \includegraphics[width=\columnwidth]{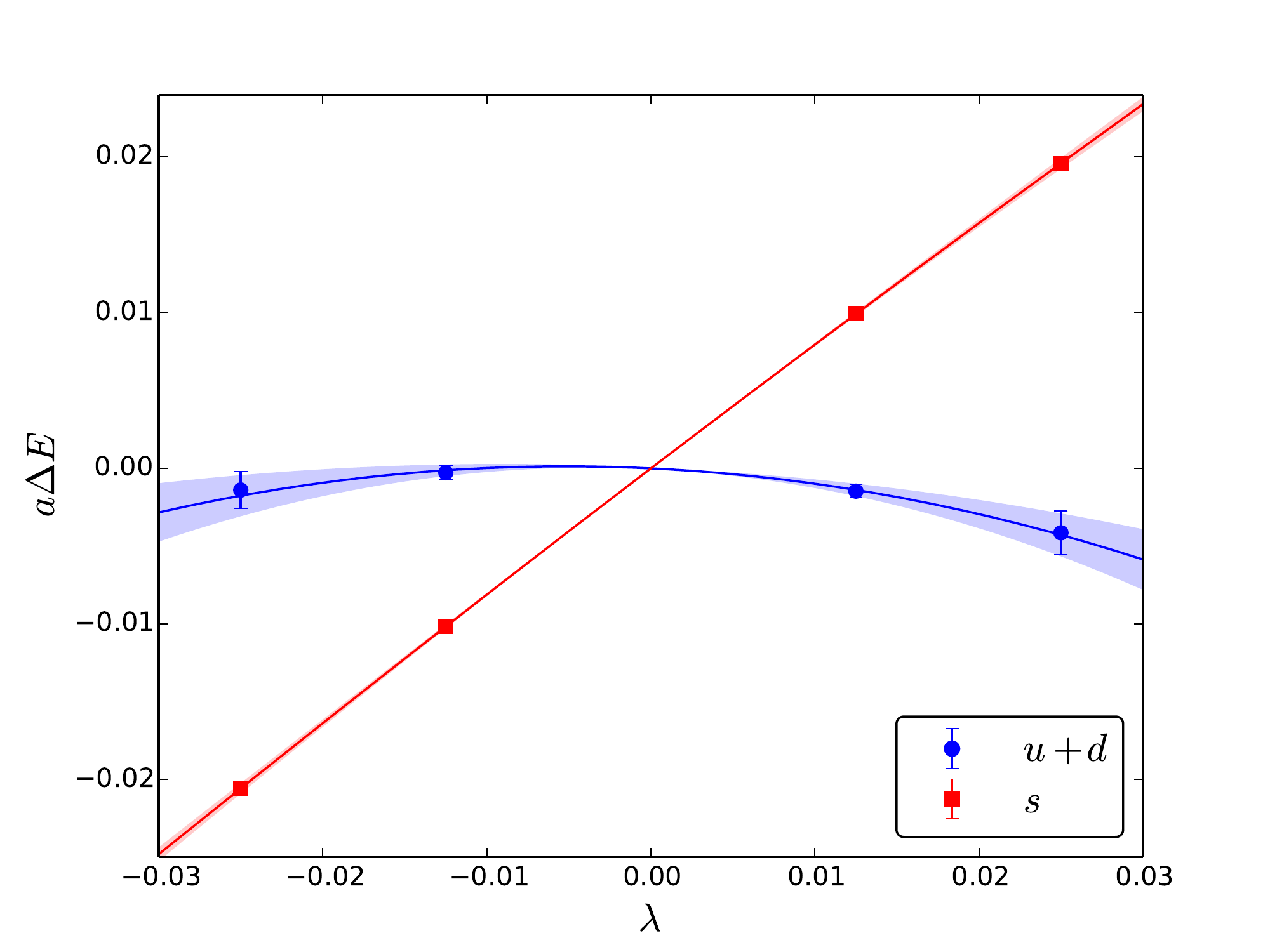}
  \caption{Energy shift of the $\Lambda$ baryon with respect to $\lambda$. $\kappa_l
    = 0.121040$, $\kappa_s = 0.120620$.}
  \label{fig:lambda8LambdaRatio}
\end{figure}
Fig.~\ref{fig:lambda8LambdaRatio} shows results for the energy shift
of the $\Lambda$ baryon on the $\kappa_l=0.121040$, $\kappa_s =
0.120620$ ensemble for two values of $\lambda = 0.0125, 0.025$. 
The strong, highly constrained positive slope for the strange quark axial charge is consistent with the common expectation
that the heavier strange quark carries the dominant spin fraction.
 Conversely, the small negative light quark contribution is
more weakly constrained, subject predominantly only to quadratic
effects.

In order to make a comparison with existing results in the literature,
we make use of the preliminary results for $Z_A$ given in
Eq.~(\ref{eq:axialRenormalisation}). For the $\Lambda$ baryon at the
lighter pion mass of around 360 MeV, we have for light and strange
connected contributions,
\begin{align}
  {\Delta u}_\text{conn.} = {\Delta d}_\text{conn.} & = -0.043(14) \, \\ 
  {\Delta s}_\text{conn.} & = 0.683(18) \, .
\end{align}
Very few other lattice calculations of these quantities have been
performed, the only example being in \cite{Gockeler:2002} from a
chiral extrapolation of quenched calculations at pion masses of around
600 MeV and upwards,
\begin{align}
   {\Delta u}_\text{conn.} = {\Delta d}_\text{conn.} & = -0.02(4) \, ,
   \\ {\Delta s}_\text{conn.} & = 0.68(4)\, ,
\end{align}
which are in good agreement.

\subsection{Decuplet Baryons $(J = \frac{3}{2})$}

For the decuplet baryons, we make use of the interpolating operator
\begin{align}\label{eq:deltaOperator}
  \mathcal{O}_\text{decuplet} = \epsilon_{abc} \frac{1}{\sqrt{3}} [ 2 & \left( {q_1}_a^T C \gamma_\pm {q_2}_b \right) {q_1}_c \\ \notag
    & + \left({q_1}_a^T C \gamma_\pm {q_1}_b \right) {q_2}_c ],
\end{align}
where we define $\gamma_\pm$ (in Euclidean space) as
\begin{equation}
  \gamma_\pm = - i \frac{1}{2} \left( \gamma_1 \pm i \gamma_2 \right).
\end{equation}
Analagously to the case of the octet baryons, appropriate quark flavours are
substituted into Eq.~(\ref{eq:deltaOperator}) to access all decuplet
states. We again make use of the parity and spin-projection operators in
Eq.~(\ref{eq:parityProjectionOperators}) and
Eq.~(\ref{eq:spinProjectionOperators}). However we must take care when
analysing the $m = \pm \frac{1}{2}$ states.  Using the $\gamma_\pm$
matrices, we create diquarks with $J = 1$ and $m = \pm 1$. The
$\Gamma_{\pm}$ operator projects out the spin of the single quark to
$m = \pm \frac{1}{2}$ as before. For the $m = \pm \frac{3}{2}$
baryon states, there is no problem combining the diquark and single
quark, since
\begin{align}
  \left| 1 \; + 1 \right\rangle \left| \frac{1}{2} \; + \frac{1}{2}
  \right\rangle & = \left| \frac{3}{2} \; + \frac{3}{2} \right\rangle,
  \\ \left| 1 \; - 1 \right\rangle \left| \frac{1}{2} \; - \frac{1}{2}
  \right\rangle & = \left| \frac{3}{2} \; - \frac{3}{2} \right\rangle.
\end{align}
However, when we create the $m = \pm \frac{1}{2}$ states, we create
a mixture of $J = \frac{3}{2}$ and $J = \frac{1}{2}$ states,
\begin{align}
  \left| 1 \; + 1 \right\rangle \left| \frac{1}{2} \; - \frac{1}{2}
  \right\rangle & = \sqrt{\frac{1}{3}} \left| \frac{3}{2} \; +
  \frac{1}{2} \right\rangle + \sqrt{\frac{2}{3}} \left| \frac{1}{2} \;
  + \frac{1}{2} \right\rangle, \\ \left| 1 \; - 1 \right\rangle \left|
  \frac{1}{2} \; + \frac{1}{2} \right\rangle & = \sqrt{\frac{1}{3}}
  \left| \frac{3}{2} \; - \frac{1}{2} \right\rangle -
  \sqrt{\frac{2}{3}} \left| \frac{1}{2} \; - \frac{1}{2}
  \right\rangle.
\end{align}

In principle, it is possible to project onto definite $J =
\frac{3}{2},\frac{1}{2}$ states (see \cite{Zanotti:2003}).  However,
we note that the $J = \frac{1}{2}$, $\Delta (1750)$ state has a higher
mass than the $J = \frac{3}{2}$, $\Delta (1232)$ state, and so we
expect the $\Delta (1232)$ to saturate the ground state at large Euclidean
time. Although we may expect to see slightly more excited-state
contamination than in the $m = \pm \frac{3}{2}$ cases.

\begin{figure}
  \includegraphics[width=\columnwidth]{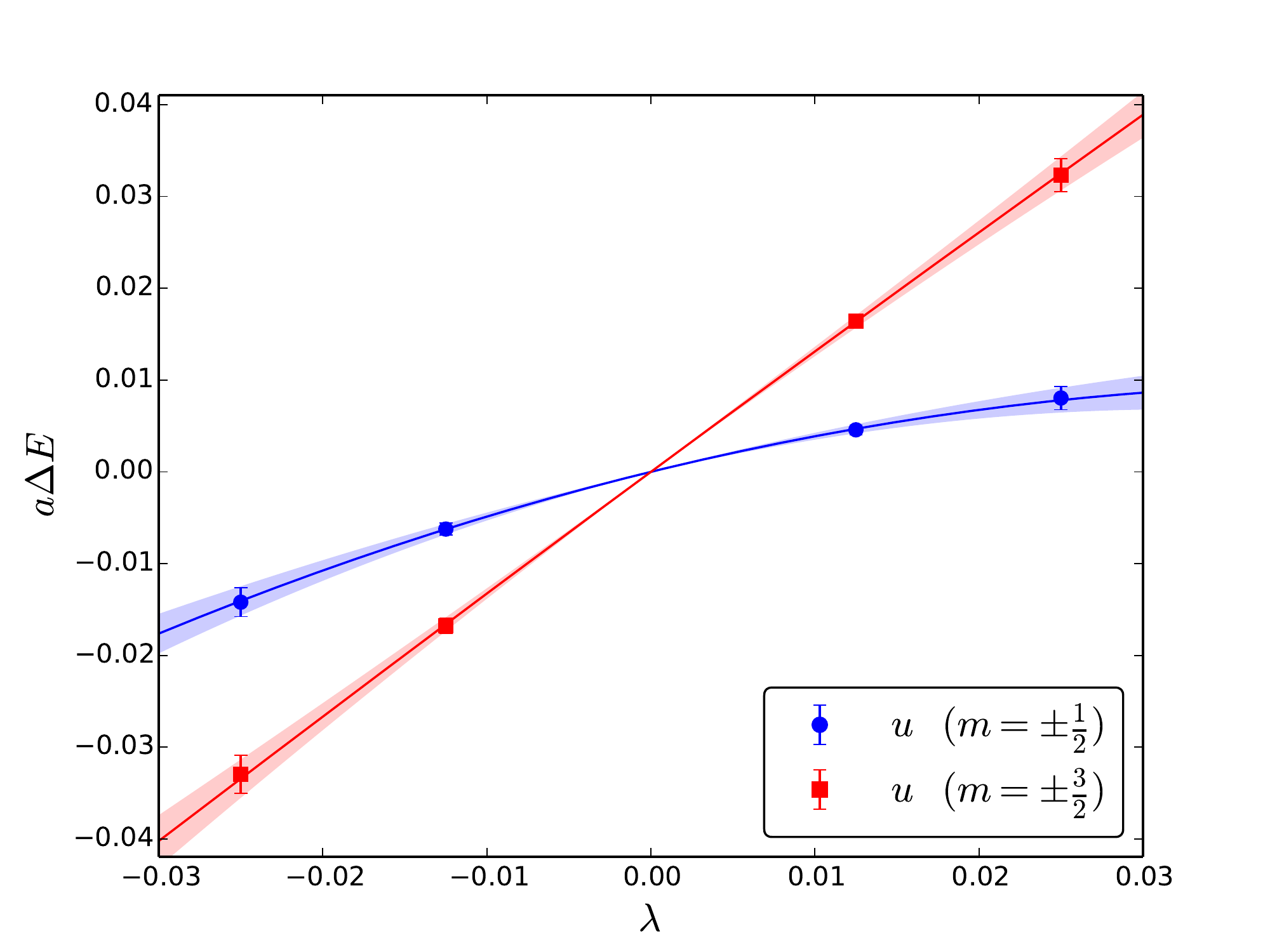}
  \caption{Energy shift of $\Delta^+$ with respect to $\lambda$
    parameter on the $\kappa_l = 0.121040$, $\kappa_s = 0.120620$
    ensemble. Only results for the up quark are shown (Results for the
    down quark differ by a factor of a half).}
  \label{fig:deltaLambdaRatio}
\end{figure}
Fig.~\ref{fig:deltaLambdaRatio} shows results for the energy shift of
the $\Delta$ baryon with $\lambda$. Only results for the up quark are
shown, since the spin-contribution for the down quark differs by a factor of a
half in the isospin-symmetric limit. In contrast to the nucleon, all quarks in the $\Delta$ have positive
contributions.

Table~\ref{tab:decupletResults} summarises results for the decuplet
baryons. Results at the heavier pion mass for the $m = \pm
\frac{1}{2}$ states are unavailable, as the code to calculate these
was not implemented at the time of those initial runs. Note that the distinction between the two different 
quark flavour spin contributions is a result of the form of the interpolating operator. So for example, the overall
strange connected contribution to the $\Omega$ baryon, ${\Delta s}^{\frac{3}{2} m} =
{\Delta q_1}^{\frac{3}{2} m} + {\Delta q_2}^{\frac{3}{2} m}$.
\begin{table}
  \begin{tabular}{ c|c c c c}
    \hline
    \hline
    $\kappa_l$ & \multicolumn{2}{c}{0.120900} & \multicolumn{2}{c}{0.121040} \\
    $\kappa_s$ & \multicolumn{2}{c}{0.120900} & \multicolumn{2}{c}{0.120620} \\
    \hline
    & ${\Delta q_1}_\text{conn.}^\text{latt.}$ & ${\Delta q_2}_\text{conn.}^\text{latt.}$ & ${\Delta q_1}_\text{conn.}^\text{latt.}$ & ${\Delta q_2}_\text{conn.}^\text{latt.}$ \\
    \hline
    $\Delta (m = \pm \frac{3}{2}) $ & 1.364(29) & 0.682(15) & 1.319(48) & 0.660(24) \\
    $\Sigma^* (m = \pm \frac{3}{2}) $ & 1.364(29) & 0.682(15) & 1.310(43) & 0.727(11) \\
    $\Xi^* (m = \pm \frac{3}{2}) $ & 1.364(29) & 0.682(15) & 1.448(19) & 0.654(20) \\
    $\Omega (m = \pm \frac{3}{2}) $ & 1.364(29) & 0.682(15) & 1.437(16) & 0.718(8) \\
    $\Delta (m = \pm \frac{1}{2}) $ & - & - & 0.437(36) & 0.215(18) \\
    $\Sigma^* (m = \pm \frac{1}{2}) $ & - & - & 0.441(31) & 0.244(9) \\
    $\Xi^* (m = \pm \frac{1}{2}) $ & - & - & 0.506(14) & 0.215(14) \\
    $\Omega (m =\pm \frac{1}{2}) $ & - & -& 0.504(12) & 0.248(6) \\
    \hline
    \hline
  \end{tabular}
  \caption{Table of connected spin contributions for the baryon
    decuplet. $q_1$ and $q_2$ are as they appear in the interpolating
    operator (\ref{eq:deltaOperator}).}
  \label{tab:decupletResults}
\end{table}

Similarly to the results for the octet baryons in
Table~\ref{tab:octetResults}, we see the effect of the changing quark
masses on the axial charges. We observe the same pattern of environmental sensitivity
as was evident when comparing the nucleon and $\Xi$ away from the SU(3)
symmetric point; the heavier strange quark lowers the light quark
contribution. 
For the zeroth moment \cite{Jaffe:1988up},
\begin{equation}
  {\Delta q}^{\frac{3}{2} \frac{3}{2}} =  3 {\Delta q}^{\frac{3}{2} \frac{1}{2}} .
\end{equation}
Comparing results for the $m = \pm \frac{1}{2}, \pm \frac{3}{2}$ 
states in Table~\ref{tab:decupletResults}, we see broad agreement
with the sum rule.

Using the preliminary renormalisation factor in Eq.~(\ref{eq:axialRenormalisation}), we have for the $\Delta$ baryon at the lighter pion mass of around 360 MeV,
\begin{equation}
  {\Delta u}_\text{conn.}^{\frac{3}{2} \frac{3}{2}} + {\Delta d}_\text{conn.}^{\frac{3}{2} \frac{3}{2}} = 1.682(61) \, ,
\end{equation}
which compares well to results from \cite{Alexandrou:2013b} at a pion
mass of 297(5) MeV,
\begin{equation}
  {\Delta u}_\text{conn.}^{\frac{3}{2} \frac{3}{2}} + {\Delta d}_\text{conn.}^{\frac{3}{2} \frac{3}{2}} = 1.81(11) \, .
\end{equation}

\subsection{Vector Mesons $(J = 1)$}

For the vector mesons, we make use of the interpolating operator
\begin{equation}\label{eq:rhoInterpolator}
  \mathcal{O}_\text{vector} = \overline{q}_2 \gamma_\pm q_1,
\end{equation}
where again appropriate quark flavours are substituted to access the
different meson states.

\begin{figure}
  \includegraphics[width=\columnwidth]{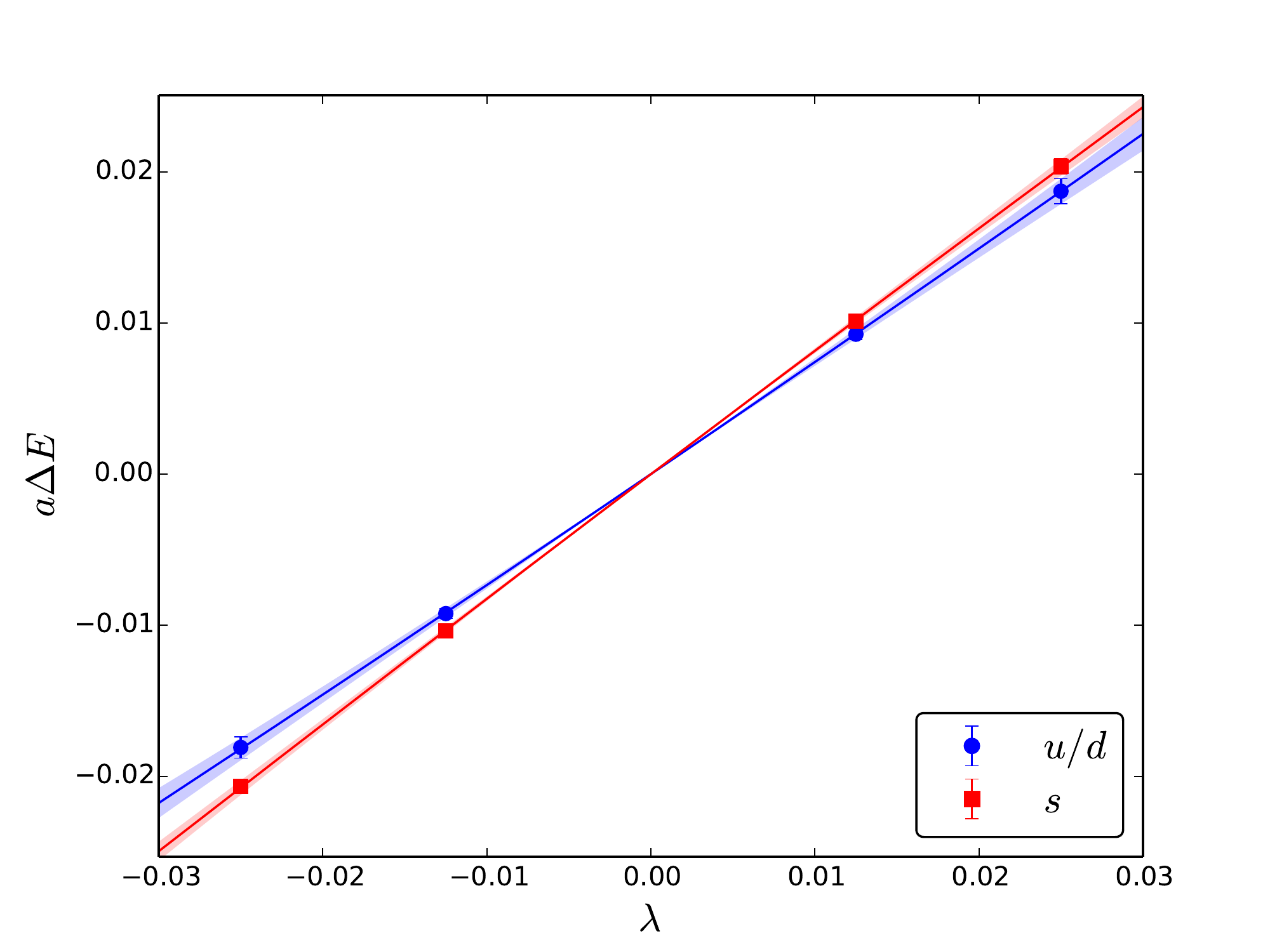}
  \caption{Energy shift of $K^*$ with respect to $\lambda$ parameter
    on $\kappa_l = 0.121040$, $\kappa_s = 0.120620$ ensemble.}
  \label{fig:kstarLambdaRatio}
\end{figure}
Fig.~\ref{fig:kstarLambdaRatio} shows results for the $K^*$ meson. We
see that both quark flavours have positive contributions, and observe
the slightly larger contribution from the strange quark.
Table~\ref{tab:vectorMesonResults} summarises results for the vector
mesons.
\begin{table}
  \begin{tabular}{c |c c c c}
    \hline
    \hline
    $\kappa_l$ & \multicolumn{2}{c}{0.120900} & \multicolumn{2}{c}{0.121040} \\
    $\kappa_s$ & \multicolumn{2}{c}{0.120900} & \multicolumn{2}{c}{0.120620} \\
    \hline
    & ${\Delta q_1}_\text{conn.}^\text{latt.}$ & ${\Delta q_2}_\text{conn.}^\text{latt.}$ & ${\Delta q_1}_\text{conn.}^\text{latt.}$ & ${\Delta q_2}_\text{conn.}^\text{latt.}$ \\
    \hline
    $\rho$ & 0.762(14) & 0.762(14) & 0.771(33) & 0.771(33) \\
    $K^*$ & 0.762(14) & 0.762(14) & 0.738(22) & 0.821(15) \\
    $\phi$ & 0.762(14) & 0.762(14) & 0.793(11) & 0.793(11) \\
    \hline
    \hline
  \end{tabular}
  \caption{Table of connected spin contributions for the vector
    mesons. $q_1$ refers to the first flavour in
    Eq.~(\ref{eq:rhoInterpolator}), and $q_2$ to the second.}
  \label{tab:vectorMesonResults}
\end{table}

We find relatively little change in the quark axial charges in the
$\rho$ at the different quark masses (consistent with results reported in
\cite{Best:1997}). We do see a similar environmental sensitivity as in
the octet and decuplet away from the SU(3) symmetric point. For
example the strange spin contribution to the $K^*$ is greater than
that for the $\phi$ due to the presence of the light quark in the
$K^*$.

For the $\rho$ meson at the lighter pion mass of around 360 MeV, we
have for the light spin contribution, using the preliminary
renormalisation in Eq.~(\ref{eq:axialRenormalisation}),
\begin{equation}
  {\Delta u}_\text{conn.}^{1 1} + {\Delta d}_\text{conn.}^{1 1} = 1.311(64) \, ,
\end{equation}
noting that the results quoted in Table~\ref{tab:vectorMesonResults}
are for each light quark individually.  Again, this calculation is
rare in the literature. Ref.~\cite{Best:1997} quotes a value, after chiral extrapolation
of quenched results, of
\begin{equation}
  {\Delta u}_\text{conn.}^{1 1} + {\Delta d}_\text{conn.}^{1 1} = 1.180(92) \, ,
\end{equation}
where we see broad agreement with our results.

\subsection{Summary}

In order to compare the relative contributions of quarks to the spin of the different hadrons, we define the quark spin fraction for a spin-$J$ hadron to be,
\begin{equation}\label{eq:spinFractionDefinition}
  \widehat{\Delta \Sigma}^J = \frac{{\Delta \Sigma}^J}{2J} .
\end{equation}
\begin{figure}
  \includegraphics[width=\columnwidth]{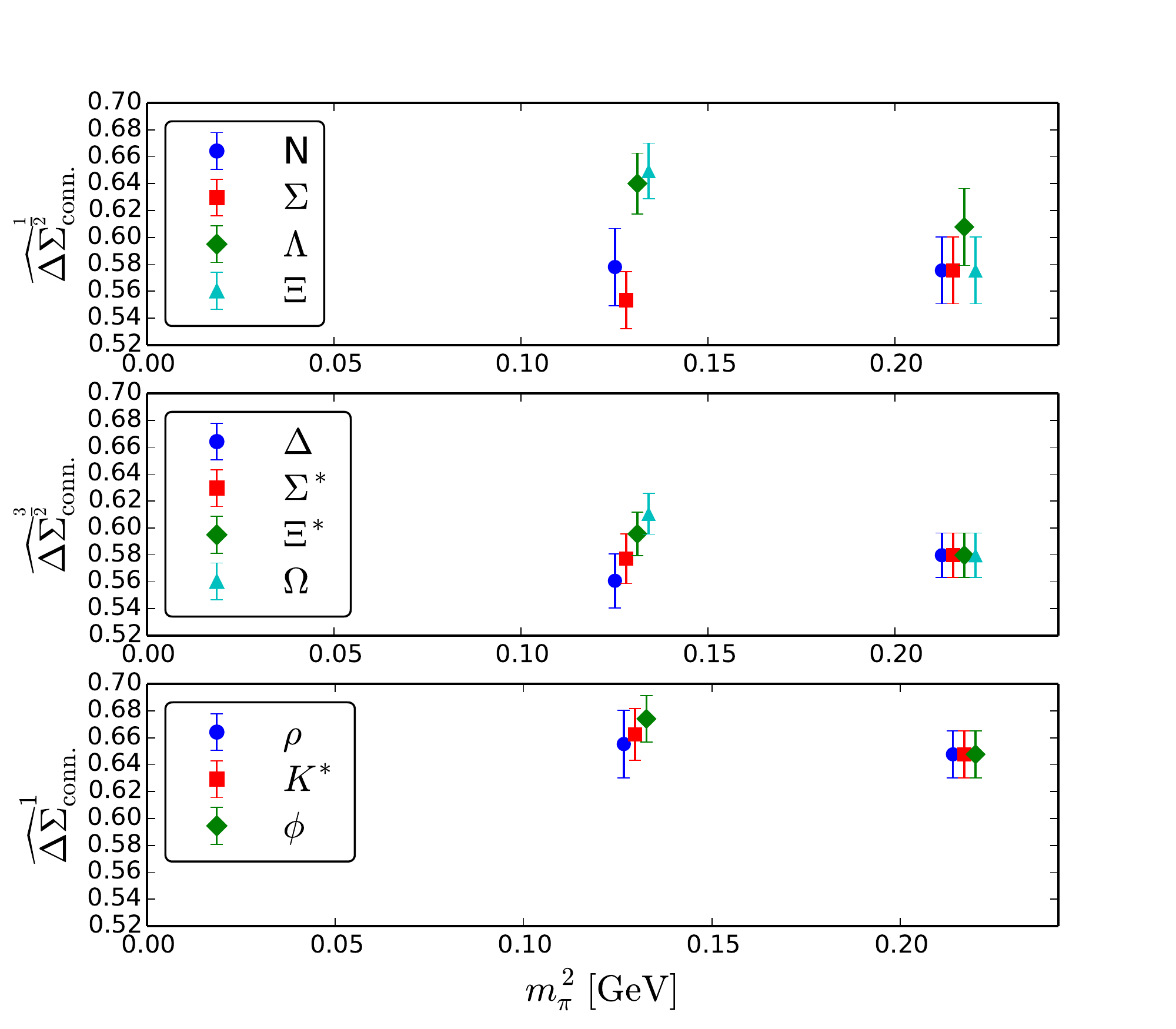}
  \caption{The quark spin fraction $\widehat{\Delta \Sigma}^J_\text{conn.}$ as defined in Eq.~(\ref{eq:spinFractionDefinition}) for all the octet and
    decuplet baryons, and vector mesons, for different pion
    masses, renormalised using Eq.~(\ref{eq:axialRenormalisation})}
  \label{fig:summaryPlot}
\end{figure}
{}Fig.~\ref{fig:summaryPlot} shows $\widehat{\Delta \Sigma}^J_\text{conn.}$ at
different pion masses for all the baryons and mesons we have examined,
renormalised using Eq.~(\ref{eq:axialRenormalisation}). Noting that the singlet quark mass is constant
along our trajectory, we see that hadrons with dominant strange quark
contributions have been shifted up with the increased strange mass,
and hadrons with dominant light quark contributions have shifted down.
We also see that the quark spin fraction for all the baryons studied here is
in the range 55-70\%.

\section{Concluding remarks}
\label{sec:fin}

We have demonstrated that the Feynman-Hellmann method is an effective
approach to calculating hadron matrix elements. 
We have demonstrated this through the determination of quark spin
contributions to hadrons.
With the statistical
improvements gained by examining hadron energy shifts, our
calculations are able to achieve comparable precision to standard
three-point function methods, with an equivalent computational investment.
It is also possible that the Feynman-Hellmann method is
less susceptible to excited-state contamination than these methods;
a current point of debate within the lattice community.

We have also shown how the Feynman-Hellmann method may be most
efficiently applied. In particular, it appears that only a couple of
different background field strengths need be realised in order to make
an accurate and precise calculation. Weaker field strengths give more
tightly constrained fit parameters, and introduce less noise to
correlation functions.

Our findings indicate that the possible application of the Feynman-Hellmann method to the
calculation of such quantities as the disconnected quark spin
contributions of hadrons is extremely promising. These simulations
will require generating separate sets of gauge field configurations,
and a significant investment of computational time. However the
possibility of accessing such matrix elements so simply and with
minimal excited state contamination is extremely promising.

\section*{Acknowledgements}
The numerical configuration generation was performed using the BQCD
lattice QCD program, \cite{Nakamura:2010},
 on the IBM BlueGeneQ using
DIRAC 2 resources (EPCC, Edinburgh, UK), the BlueGene P and Q at NIC
(J\"ulich, Germany) and the Cray XC30 at HLRN (The North-German Supercomputing Alliance).
Some of the simulations were undertaken on the NCI National Facility
in Canberra, Australia, which is supported by the Australian
Commonwealth Government.
The BlueGene codes were optimised using Bagel \cite{Boyle:2009}.
The Chroma software library \cite{Edwards:2004},
 was used in the
data analysis.
This investigation has been supported partly by the EU grants 283286
(HadronPhysics3), 227431 (Hadron Physics2) and by the Australian Research Council under grants
FT120100821, FT100100005 and DP140103067 (RDY and JMZ).

\begin{widetext}

\appendix

\section{The Feynman-Hellmann Theorem}
\label{app:fh}

Deriving the Feynman-Hellmann theorem in a
field-theoretic sense is slightly different to the standard quantum
mechanical approach found in textbooks.
There are some constructions that must
first be introduced, and we proceed by examining both two-point and
three-point correlation functions.

\subsection{Lattice Correlation Functions}\label{sec:latticeCorrelationFunctions}

We begin our discussion with the standard definitions of lattice
two-point and three-point functions. Suppose, without loss of generality, that we have chosen
$\overline{\chi}$ and $\chi$ as creation and annihilation operators for some hadron, such as the nucleon. For the Fourier transformed
two-point function we have
\begin{equation}\label{eq:twoPointFourierTransformed}
  \int  \mathrm{d}^3 x \, e^{-i \vec{k} \cdot \vec{x} } \langle \chi(\vec{x},t) \overline{\chi}(0) \rangle = \sum_n \frac{e^{-E_n(\vec{k})t}}{2 E_n(\vec{k})} | \langle \Omega | \mathcal{\chi}(0) | n,\vec{k} \rangle |^2 \, ,
\end{equation}
where the right-hand side includes a sum over the excited states created by the chosen operators, and $| \Omega \rangle$ denotes the vacuum.
This expression may be obtained using transfer matrix methods.
At large Euclidean times, the summation is dominated by the ground state contribution,
\begin{equation}\label{eq:twoPointReadyForLarget}
  \int  \mathrm{d}^3 \vec{x} \, e^{-i \vec{k} \cdot \vec{x} } \langle
  \chi(\vec{x},t) \overline{\chi}(0) \rangle \overset{\text{large }
    t}{\longrightarrow} \frac{e^{-E_0(\vec{k})t}}{2 E_0(\vec{k})} | \langle \Omega |
  \mathcal{\chi}(0) | 0,\vec{k} \rangle |^2\, .
\end{equation}
For the Fourier-transformed three-point function, we have
\begin{equation} \label{eq:readyForTrickPart}
  \int  \mathrm{d}^3 \vec{x} \,  \mathrm{d}^3 \vec{y} \, e^{-i \vec{k} \cdot \vec{x} }
  \langle \chi(\vec{x},t) \mathcal{O}(\vec{y},\tau) \overline{\chi}(0)
  \rangle = \sum_{n,m} \frac{e^{-E_n(\vec{k}) (t - \tau)} }{2
    E_n(\vec{k})} \frac{e^{-E_m(\vec{k}) \tau}}{2 E_m(\vec{k})}
  \ \langle \Omega | \chi(0) | n,\vec{k} \rangle \langle n,\vec{k} |
  \mathcal{O}(0) | m,\vec{k} \rangle \langle m,\vec{k} |
  \overline{\chi}(0) | \Omega \rangle\, ,
\end{equation}
where we constrain ourselves here to the special case of zero
momentum transfer between initial and final states.
Integrating both sides of this expression with respect to $\tau$, we have
\begin{align} \label{eq:readyForTrickPart2}
  \int_0^t  \mathrm{d} \tau \int  \mathrm{d}^3 \vec{x} \,  \mathrm{d}^3 \vec{y} \, e^{-i \vec{k} \cdot \vec{x} } \langle \chi(\vec{x},t) \mathcal{O}(\vec{y},\tau) \overline{\chi}(0) \rangle =
  \sum_{n,m} & \frac{e^{-E_m(\vec{k}) t} - e^{-E_n(\vec{k}) t}}{4 E_n(\vec{k}) E_m(\vec{k}) (E_n(\vec{k}) - E_m(\vec{k}))} \\ \notag
  & \times \langle \Omega | \chi(0) | n,\vec{k} \rangle \langle n,\vec{k} | \mathcal{O}(0) | m,\vec{k} \rangle \langle m,\vec{k} | \overline{\chi}(0) | \Omega \rangle\, .
\end{align}
Now consider the large $t$ behaviour of the right-hand side of this
equation.
When we expand the sums over $n$ and $m$, the dominant terms at large
$t$ will be those with the lowest values of $E_n$ and $E_m$, when $n=m=0$.
However, note that when $E_n = E_m$, the right-hand side is
ill-defined.
So we first take the limit as $E_m \to E_n$ using
l'H\^{o}pital's rule.
Dropping explicit momentum dependence momentarily, we have that
\begin{equation}
  \lim_{E_m \to E_n} \frac{e^{-E_m t} - e^{-E_n t}}{4 E_n E_m (E_n -
    E_m)} = \frac{te^{-E_n t} }{ 4 E_n^2}\, .
\end{equation}
The large $t$ behaviour of Eq.~(\ref{eq:readyForTrickPart2}) is then
given by
\begin{equation}\label{eq:threePointThing}
  \int_0^t  \mathrm{d} \tau \int  \mathrm{d}^3 \vec{x} \,  \mathrm{d}^3 \vec{y} \, e^{-i \vec{k} \cdot \vec{x} } \langle \chi(\vec{x},t) \mathcal{O}(\vec{y},\tau) \overline{\chi}(0) \rangle 
  \overset{\text{large } t}{\longrightarrow}
  \frac{te^{-E_H(\vec{k}) t} }{4 E_H^2(\vec{k})} | \langle \Omega | \chi(0) | H, \vec{k} \rangle |^2 \langle H, \vec{k} | \mathcal{O}(0) | H, \vec{k} \rangle\, ,
\end{equation}
where we denote the ground state of our hadron as $| H, \vec{k} \rangle$ and its energy to be $E_H$

\subsection{The Feynman-Hellmann Theorem}
We now proceed with a proof of
the Feynman-Hellmann theorem, expanding upon a proof presented in \cite{Horsley:2012}.
Suppose that we modify the action $S$ of our theory in some way, such
that it now depends on some parameter $\lambda$,
\begin{equation}
  S \to S(\lambda) \, .
\end{equation}
Consider the two-point correlation function discussed in the previous section.
In the path integral formalism, this correlator is given by
\begin{equation}\label{eq:readyToDifferentiate}
  \langle \chi(\vec{x},t) \overline{\chi}(0) \rangle_\lambda =
  \frac{1}{Z(\lambda)} \int \mathcal{D} A \mathcal{D} \psi
  \mathcal{D} \overline{\psi} \, \chi(\vec{x},t) \overline{\chi}(0)
  e^{-S(\lambda)}\, ,
\end{equation}
where the subscript $\lambda$ indicates that the correlator is to be
evaluated with respect to the modified action, and we note that the partition function is now also a function of
$\lambda$,
\begin{equation}
  Z(\lambda) = \int \mathcal{D} A \mathcal{D} \psi \mathcal{D}
  \overline{\psi} \, e^{-S(\lambda)}\, .
\end{equation}
Taking the derivative with respect to $\lambda$ of both sides of
Eq.~(\ref{eq:readyToDifferentiate}), it is straightforward to show that
\begin{equation}\label{eq:readyToFourierTransform}
  \frac{\partial}{\partial \lambda} \langle \chi(\vec{x},t)
  \overline{\chi}(0) \rangle_\lambda = \left\langle \frac{\partial
    S(\lambda)}{\partial \lambda} \right\rangle_\lambda \langle
  \chi(\vec{x},t) \overline{\chi}(0) \rangle_\lambda - \left\langle
  \chi(\vec{x},t) \frac{\partial S(\lambda)}{\partial \lambda}
  \overline{\chi}(0) \right\rangle_\lambda \, ,
\end{equation}
noting that angular brackets here denote expectation values as given in the path-integral formalism, analagous to Eq.~(\ref{eq:readyToDifferentiate}),
\begin{equation}
  \langle \mathcal{O} \rangle = \frac{1}{Z} \int \mathcal{D} A \mathcal{D} \psi \mathcal{D} \overline{\psi} \, \mathcal{O} \, e^{-S} \, .
\end{equation}
Fourier transforming both sides of Eq.~(\ref{eq:readyToFourierTransform}) and re-arranging terms, we obtain the expression
\begin{equation}\label{eq:dealWithSidesSeparately}
   \left\{ \frac{\partial}{\partial \lambda} - \left\langle
   \frac{\partial S(\lambda)}{\partial \lambda} \right\rangle_\lambda
   \right\} \int \mathrm{d}^3 \vec{x} \, e^{-i \vec{k} \cdot \vec{x} } \langle
   \chi(\vec{x},t) \overline{\chi}(0) \rangle_\lambda = - \int \mathrm{d}^3
   \vec{x} \, e^{-i \vec{k} \cdot \vec{x} } \left\langle
   \chi(\vec{x},t) \frac{\partial S(\lambda)}{\partial \lambda}
   \overline{\chi}(0) \right\rangle_\lambda \, .
\end{equation}
Consider the first term on the left-hand side of this expression.
We have a derivative with respect to $\lambda$ of the two-point
correlator from Appendix~\ref{sec:latticeCorrelationFunctions}.
Since our action now depends on the parameter $\lambda$, we
have
\begin{equation}\label{eq:readyForDerivative}
  \int \mathrm{d}^3 x \, e^{-i \vec{k} \cdot \vec{x} } \langle \chi(\vec{x},t) \overline{\chi}(0) \rangle_\lambda 
  = \sum_n \frac{e^{-E_n(\vec{k},\lambda)t}}{2 E_n(\vec{k},\lambda)} | \langle \Omega | \mathcal{\chi}(0) | n,\vec{k} \rangle_\lambda |^2 \, ,
\end{equation}
noting that the energy eigenvalues and amplitudes both depend on
$\lambda$.
In deriving this expression, we required that the vacuum
state has zero energy.
We note that in modifying our action, we may have shifted our vacuum
energy to a non-zero value (for instance, if our
  modification to the action took the form of the operator $\lambda
  \overline{q} q$ for some parameter $\lambda$).
However, we will assume that this is not the case, as the
modifications we make to the action in the main body of the paper,
namely the inclusion of the axial operator $\overline{q} i \gamma_\mu \gamma_5 q$, does not shift the vacuum energy.
We can calculate the derivative with respect to $\lambda$ of Eq.~(\ref{eq:readyForDerivative}),
\begin{equation}
  \frac{\partial}{\partial \lambda} \int  \mathrm{d}^3 x \, e^{-i \vec{k} \cdot \vec{x} } \langle \chi(\vec{x},t) \overline{\chi}(0) \rangle_\lambda 
  = \sum_n  \frac{e^{-E_n(\vec{k},\lambda) t}}{2 E_n(\vec{k},\lambda)} \left\{ 
  - \left( t +  \frac{1}{E_n(\vec{k},\lambda)} \right) \frac{\partial E_n(\vec{k},\lambda)} {\partial \lambda} 
  + \frac{\partial}{\partial \lambda} \right\}  | \langle \Omega | \mathcal{\chi}(0) | n,\vec{k} \rangle_\lambda |^2  \, .
\end{equation}
At large Euclidean times, the lowest energy state in the summation above will dominate the summation, and the term with linear time dependence will dominate 
the second and third terms. Hence we have
\begin{equation}\label{eq:firstSub}
  \frac{\partial}{\partial \lambda} \int  \mathrm{d}^3 x \, e^{-i \vec{k} \cdot
    \vec{x} } \langle \chi(\vec{x},t) \overline{\chi}(0) \rangle_\lambda
  \overset{\text{large } t}{\longrightarrow} - \frac{\partial
    E_H(\vec{k},\lambda)} {\partial \lambda} \frac{t
    e^{-E_H(\vec{k},\lambda) t}}{2 E_H(\vec{k},\lambda)} | \langle
  \Omega | \mathcal{\chi}(0) | H, \vec{k} \rangle_\lambda |^2\, .
\end{equation}
Next, consider the second term on the left-hand side of
Eq.~(\ref{eq:dealWithSidesSeparately}),
\begin{equation}\label{eq:vacuumTerm}
\left\langle \frac{\partial S(\lambda)}{\partial \lambda}
\right\rangle_\lambda \int  \mathrm{d}^3 \vec{x} \, e^{-i \vec{k} \cdot \vec{x}
} \langle \chi(\vec{x},t) \overline{\chi}(0) \rangle_\lambda\, .
\end{equation}
The very first quantity is just a vacuum expectation value, and assuming that the modification of $S$ does not carry vacuum quantum
numbers (to leading order in $\lambda$), this contribution will
vanish.
Finally, consider the term on the right-hand side of
Eq.~(\ref{eq:dealWithSidesSeparately}),
\begin{equation}
  \int  \mathrm{d}^3 \vec{x} \, e^{-i \vec{k} \cdot \vec{x} } \left\langle
  \chi(\vec{x},t) \frac{\partial S(\lambda)}{\partial \lambda}
  \overline{\chi}(0) \right\rangle_\lambda\, .
\end{equation}
Defining the operator $\mathcal{O}$ such that
\begin{equation} \label{eq:ODefinition}
  \int \mathrm{d} \tau \int  \mathrm{d}^3 \vec{y} \, \mathcal{O}(\vec{y},\tau) =
  \frac{\partial S(\lambda)}{\partial \lambda}\, ,
\end{equation}
we have exactly the three-point correlator described by
Eq.~(\ref{eq:threePointThing}), noting however that the energies and
amplitudes now have explicit $\lambda$ dependence. 
We also point out that while the implementation of the operator is made across the whole lattice, the correlation function will only receive a significant contribution between $0$ and $t$. Hence, we restrict the $\tau$ integration to this domain, and have
So
\begin{equation} \label{eq:secondSub}
  \int_0^t  \mathrm{d} \tau \int  \mathrm{d}^3 \vec{x} \,  \mathrm{d}^3 \vec{y} \, e^{-i \vec{k} \cdot \vec{x}} \langle \chi(\vec{x},t) \mathcal{O}(\vec{y},\tau) \overline{\chi}(0) \rangle_\lambda \overset{\text{large } t}{\longrightarrow}
  \frac{te^{-E_H(\vec{k},\lambda) t} }{4 E_H^2(\vec{k},\lambda)} | \langle \Omega | \chi(0) | H, \vec{k} \rangle_\lambda |^2 \langle H, \vec{k} | \mathcal{O}(0) | H, \vec{k} \rangle_\lambda\, .
\end{equation}
As above, we again assume the modification to the action
has not shifted the vacuum energy.
So starting from Eq.~(\ref{eq:dealWithSidesSeparately}) and taking the
behaviour at large $t$ on both sides, substituting in
Eq.~(\ref{eq:firstSub}) and Eq.~(\ref{eq:secondSub}) we have
\begin{equation}
  - \frac{\partial E(\vec{k},\lambda)} {\partial \lambda} \frac{t e^{-E_H(\vec{k},\lambda) t}}{2 E_H(\vec{k},\lambda)} | \langle \Omega | \mathcal{\chi}(0) | H, \vec{k} \rangle_\lambda |^2 \notag = 
  - \frac{te^{-E_H(\vec{k},\lambda) t} }{4 E_H^2(\vec{k},\lambda)} | \langle \Omega | \chi(0) | H, \vec{k} \rangle_\lambda |^2 \langle H, \vec{k} | \mathcal{O}(0) | H, \vec{k} \rangle_\lambda .
\end{equation}
Cancelling various factors, we obtain
\begin{equation}
  \frac{\partial E_H(\vec{k},\lambda)} {\partial \lambda} =
  \frac{1}{2 E_H(\vec{k},\lambda)} \langle H, \vec{k} | \mathcal{O}(0) | H, \vec{k} \rangle_\lambda .
\end{equation}

We can generalise this result to any hadron for which we can choose suitable interpolating operators.
Additionally, the origin $0$ was taken only as a convenient reference point.
So in general for any hadron state $| H \rangle$, we have
\begin{equation}
  \frac{\partial E_H(\lambda)} {\partial \lambda} =
  \frac{1}{2 E_H(\lambda)} \langle H | \mathcal{O} | H \rangle_\lambda .
\end{equation}

This is our expression for the Feynman-Hellmann theorem in the context
of field theory.

\end{widetext}

\bibliography{spinFHrefs}

\begin{thebibliography}{40}
\expandafter\ifx\csname natexlab\endcsname\relax\def\natexlab#1{#1}\fi
\expandafter\ifx\csname bibnamefont\endcsname\relax
  \def\bibnamefont#1{#1}\fi
\expandafter\ifx\csname bibfnamefont\endcsname\relax
  \def\bibfnamefont#1{#1}\fi
\expandafter\ifx\csname citenamefont\endcsname\relax
  \def\citenamefont#1{#1}\fi
\expandafter\ifx\csname url\endcsname\relax
  \def\url#1{\texttt{#1}}\fi
\expandafter\ifx\csname urlprefix\endcsname\relax\def\urlprefix{URL }\fi
\providecommand{\bibinfo}[2]{#2}
\providecommand{\eprint}[2][]{\url{#2}}

\bibitem[{\citenamefont{Alexakhin et~al.}(2007)}]{Alexakhin:2006oza}
\bibinfo{author}{\bibfnamefont{V.~Y.} \bibnamefont{Alexakhin}}
  \bibnamefont{et~al.} (\bibinfo{collaboration}{COMPASS Collaboration}),
  \bibinfo{journal}{Phys.~Lett.} \textbf{\bibinfo{volume}{B647}},
  \bibinfo{pages}{8} (\bibinfo{year}{2007}),
\eprint{hep-ex/0609038}.
%%CITATION = HEP-EX/0609038;%%.

\bibitem[{\citenamefont{Anselmino et~al.}(1995)\citenamefont{Anselmino,
  Efremov, and Leader}}]{Anselmino:1994gn}
\bibinfo{author}{\bibfnamefont{M.}~\bibnamefont{Anselmino}},
  \bibinfo{author}{\bibfnamefont{A.}~\bibnamefont{Efremov}}, \bibnamefont{and}
  \bibinfo{author}{\bibfnamefont{E.}~\bibnamefont{Leader}},
  \bibinfo{journal}{Phys.~Rept.} \textbf{\bibinfo{volume}{261}},
  \bibinfo{pages}{1} (\bibinfo{year}{1995}),
\eprint{hep-ph/9501369}.
%%CITATION = HEP-PH/9501369;%%.

\bibitem[{\citenamefont{Filippone and Ji}(2001)}]{Filippone:2001ux}
\bibinfo{author}{\bibfnamefont{B.~W.} \bibnamefont{Filippone}}
  \bibnamefont{and} \bibinfo{author}{\bibfnamefont{X.-D.} \bibnamefont{Ji}},
  \bibinfo{journal}{Adv.~Nucl.~Phys.} \textbf{\bibinfo{volume}{26}},
  \bibinfo{pages}{1} (\bibinfo{year}{2001}),
\eprint{hep-ph/0101224}.
%%CITATION = HEP-PH/0101224;%%.

\bibitem[{\citenamefont{Bass}(2005)}]{Bass:2004xa}
\bibinfo{author}{\bibfnamefont{S.~D.} \bibnamefont{Bass}},
  \bibinfo{journal}{Rev.~Mod.~Phys.} \textbf{\bibinfo{volume}{77}},
  \bibinfo{pages}{1257} (\bibinfo{year}{2005}),
\eprint{hep-ph/0411005}.
%%CITATION = HEP-PH/0411005;%%.

\bibitem[{\citenamefont{Aidala et~al.}(2013)\citenamefont{Aidala, Bass, Hasch,
  and Mallot}}]{Aidala:2012mv}
\bibinfo{author}{\bibfnamefont{C.~A.} \bibnamefont{Aidala}},
  \bibinfo{author}{\bibfnamefont{S.~D.} \bibnamefont{Bass}},
  \bibinfo{author}{\bibfnamefont{D.}~\bibnamefont{Hasch}}, \bibnamefont{and}
  \bibinfo{author}{\bibfnamefont{G.~K.} \bibnamefont{Mallot}},
  \bibinfo{journal}{Rev.~Mod.~Phys.} \textbf{\bibinfo{volume}{85}},
  \bibinfo{pages}{655} (\bibinfo{year}{2013}),
\eprint{1209.2803}.
%%CITATION = ARXIV:1209.2803;%%.

\bibitem[{\citenamefont{Alexandrou
  et~al.}(2013{\natexlab{a}})}]{Alexandrou:2013joa}
\bibinfo{author}{\bibfnamefont{C.}~\bibnamefont{Alexandrou}}
  \bibnamefont{et~al.}, \bibinfo{journal}{Phys.~Rev.}
  \textbf{\bibinfo{volume}{D88}}, \bibinfo{pages}{014509}
  (\bibinfo{year}{2013}{\natexlab{a}}),
\eprint{1303.5979}.
%%CITATION = ARXIV:1303.5979;%%.

\bibitem[{\citenamefont{Bratt et~al.}(2010)}]{Bratt:2010jn}
\bibinfo{author}{\bibfnamefont{J.~D.} \bibnamefont{Bratt}} \bibnamefont{et~al.}
  (\bibinfo{collaboration}{LHPC Collaboration}), \bibinfo{journal}{Phys.~Rev.}
  \textbf{\bibinfo{volume}{D82}}, \bibinfo{pages}{094502}
  (\bibinfo{year}{2010}),
\eprint{1001.3620}.
%%CITATION = ARXIV:1001.3620;%%.

\bibitem[{\citenamefont{Sternbeck et~al.}(2011)}]{Sternbeck:2012rw}
\bibinfo{author}{\bibfnamefont{A.}~\bibnamefont{Sternbeck}}
  \bibnamefont{et~al.}, \bibinfo{journal}{PoS}
  \textbf{\bibinfo{volume}{LATTICE2011}}, \bibinfo{pages}{177}
  (\bibinfo{year}{2011}),
\eprint{1203.6579}.
%%CITATION = ARXIV:1203.6579;%%.

\bibitem[{\citenamefont{Syritsyn et~al.}(2011)}]{Syritsyn:2011vk}
\bibinfo{author}{\bibfnamefont{S.~N.} \bibnamefont{Syritsyn}}
  \bibnamefont{et~al.}, \bibinfo{journal}{PoS}
  \textbf{\bibinfo{volume}{LATTICE2011}}, \bibinfo{pages}{178}
  (\bibinfo{year}{2011}),
\eprint{1111.0718}.
%%CITATION = ARXIV:1111.0718;%%.

\bibitem[{\citenamefont{H{\"a}gler}(2010)}]{Hagler:2009ni}
\bibinfo{author}{\bibfnamefont{P.}~\bibnamefont{H{\"a}gler}},
  \bibinfo{journal}{Phys.~Rept.} \textbf{\bibinfo{volume}{490}},
  \bibinfo{pages}{49} (\bibinfo{year}{2010}),
\eprint{0912.5483}.
%%CITATION = ARXIV:0912.5483;%%.

\bibitem[{\citenamefont{Bali et~al.}(2012)}]{QCDSF:2011aa}
\bibinfo{author}{\bibfnamefont{G.~S.} \bibnamefont{Bali}} \bibnamefont{et~al.}
  (\bibinfo{collaboration}{QCDSF Collaboration}),
  \bibinfo{journal}{Phys.~Rev.~Lett.} \textbf{\bibinfo{volume}{108}},
  \bibinfo{pages}{222001} (\bibinfo{year}{2012}),
\eprint{1112.3354}.
%%CITATION = ARXIV:1112.3354;%%.

\bibitem[{\citenamefont{Deka et~al.}(2013)}]{Deka:2013zha}
\bibinfo{author}{\bibfnamefont{M.}~\bibnamefont{Deka}} \bibnamefont{et~al.}
  (\bibinfo{year}{2013}),
\eprint{1312.4816}.
%%CITATION = ARXIV:1312.4816;%%.

\bibitem[{\citenamefont{Detmold}(2005)}]{Detmold:2004kw}
\bibinfo{author}{\bibfnamefont{W.}~\bibnamefont{Detmold}},
  \bibinfo{journal}{Phys.~Rev.} \textbf{\bibinfo{volume}{D71}},
  \bibinfo{pages}{054506} (\bibinfo{year}{2005}),
\eprint{hep-lat/0410011}.
%%CITATION = HEP-LAT/0410011;%%.

\bibitem[{\citenamefont{Horsley et~al.}(2012)}]{Horsley:2012}
\bibinfo{author}{\bibfnamefont{R.}~\bibnamefont{Horsley}} \bibnamefont{et~al.}
  (\bibinfo{collaboration}{QCDSF/UKQCD Collaborations}),
  \bibinfo{journal}{Phys.~Lett.} \textbf{\bibinfo{volume}{B714}},
  \bibinfo{pages}{312} (\bibinfo{year}{2012}),
\eprint{1205.6410}.
%%CITATION = ARXIV:1205.6410;%%.

\bibitem[{\citenamefont{Alexandrou
  et~al.}(2013{\natexlab{b}})}]{Alexandrou:2013tfa}
\bibinfo{author}{\bibfnamefont{C.}~\bibnamefont{Alexandrou}}
  \bibnamefont{et~al.}, \bibinfo{journal}{PoS LATTICE}
  \textbf{\bibinfo{volume}{2013}}, \bibinfo{pages}{289}
  (\bibinfo{year}{2013}{\natexlab{b}}),
\eprint{1311.3174}.
%%CITATION = ARXIV:1311.3174;%%.

\bibitem[{\citenamefont{Owen et~al.}(2013)}]{Owen:2012ts}
\bibinfo{author}{\bibfnamefont{B.~J.} \bibnamefont{Owen}} \bibnamefont{et~al.},
  \bibinfo{journal}{Phys.~Lett.} \textbf{\bibinfo{volume}{B723}},
  \bibinfo{pages}{217} (\bibinfo{year}{2013}),
\eprint{1212.4668}.
%%CITATION = ARXIV:1212.4668;%%.

\bibitem[{\citenamefont{Capitani et~al.}(2012)}]{Capitani:2012gj}
\bibinfo{author}{\bibfnamefont{S.}~\bibnamefont{Capitani}}
  \bibnamefont{et~al.}, \bibinfo{journal}{Phys.~Rev.}
  \textbf{\bibinfo{volume}{D86}}, \bibinfo{pages}{074502}
  (\bibinfo{year}{2012}),
\eprint{1205.0180}.
%%CITATION = ARXIV:1205.0180;%%.

\bibitem[{\citenamefont{J{\"a}ger et~al.}(2013)}]{Jager:2013kha}
\bibinfo{author}{\bibfnamefont{B.}~\bibnamefont{J{\"a}ger}}
  \bibnamefont{et~al.} (\bibinfo{year}{2013}),
\eprint{1311.5804}.
%%CITATION = ARXIV:1311.5804;%%.

\bibitem[{\citenamefont{Dinter et~al.}(2011)}]{Dinter:2011sg}
\bibinfo{author}{\bibfnamefont{S.}~\bibnamefont{Dinter}} \bibnamefont{et~al.},
  \bibinfo{journal}{Phys.~Lett.} \textbf{\bibinfo{volume}{B704}},
  \bibinfo{pages}{89} (\bibinfo{year}{2011}),
\eprint{1108.1076}.
%%CITATION = ARXIV:1108.1076;%%.

\bibitem[{\citenamefont{Bhattacharya et~al.}(2013)}]{Bhattacharya:2013ehc}
\bibinfo{author}{\bibfnamefont{T.}~\bibnamefont{Bhattacharya}}
  \bibnamefont{et~al.} (\bibinfo{year}{2013}),
\eprint{1306.5435}.
%%CITATION = ARXIV:1306.5435;%%.

\bibitem[{\citenamefont{Bali et~al.}(2013)}]{Bali:2013nla}
\bibinfo{author}{\bibfnamefont{G.~S.} \bibnamefont{Bali}} \bibnamefont{et~al.}
  (\bibinfo{year}{2013}),
\eprint{1311.7041}.
%%CITATION = ARXIV:1311.7041;%%.

\bibitem[{\citenamefont{Thomas}(1984)}]{Thomas:1982kv}
\bibinfo{author}{\bibfnamefont{A.~W.} \bibnamefont{Thomas}},
  \bibinfo{journal}{Adv.~Nucl.~Phys.} \textbf{\bibinfo{volume}{13}},
  \bibinfo{pages}{1}
 (\bibinfo{year}{1984}).
%%CITATION = ANUPB,13,1;%%.

\bibitem[{\citenamefont{Miller}(1984)}]{Miller:1984em}
\bibinfo{author}{\bibfnamefont{G.~A.} \bibnamefont{Miller}},
  \bibinfo{journal}{Int.~Rev.~Nucl.~Phys.} \textbf{\bibinfo{volume}{1}},
  \bibinfo{pages}{189}
 (\bibinfo{year}{1984}).
%%CITATION = IRNPE,1,189;%%.

\bibitem[{\citenamefont{Thomas et~al.}(1981)\citenamefont{Thomas, Theberge, and
  Miller}}]{Thomas:1981vc}
\bibinfo{author}{\bibfnamefont{A.~W.} \bibnamefont{Thomas}},
  \bibinfo{author}{\bibfnamefont{S.}~\bibnamefont{Theberge}}, \bibnamefont{and}
  \bibinfo{author}{\bibfnamefont{G.~A.} \bibnamefont{Miller}},
  \bibinfo{journal}{Phys.~Rev.} \textbf{\bibinfo{volume}{D24}},
  \bibinfo{pages}{216}
 (\bibinfo{year}{1981}).
%%CITATION = PHRVA,D24,216;%%.

\bibitem[{\citenamefont{Shanahan et~al.}(2013)}]{Shanahan:2013apa}
\bibinfo{author}{\bibfnamefont{P.}~\bibnamefont{Shanahan}}
  \bibnamefont{et~al.}, \bibinfo{journal}{Phys.~Rev.~Lett.}
  \textbf{\bibinfo{volume}{110}}, \bibinfo{pages}{202001}
  (\bibinfo{year}{2013}),
\eprint{1302.6300}.
%%CITATION = ARXIV:1302.6300;%%.

\bibitem[{\citenamefont{Jaffe and Manohar}(1989)}]{Jaffe:1988up}
\bibinfo{author}{\bibfnamefont{R.~L.} \bibnamefont{Jaffe}} \bibnamefont{and}
  \bibinfo{author}{\bibfnamefont{A.}~\bibnamefont{Manohar}},
  \bibinfo{journal}{Nucl.~Phys.} \textbf{\bibinfo{volume}{B321}},
  \bibinfo{pages}{343}
 (\bibinfo{year}{1989}).
%%CITATION = NUPHA,B321,343;%%.

\bibitem[{\citenamefont{Horsley et~al.}(2013)}]{Horsley:2013}
\bibinfo{author}{\bibfnamefont{R.}~\bibnamefont{Horsley}} \bibnamefont{et~al.},
  \bibinfo{journal}{PoS} \textbf{\bibinfo{volume}{LATTICE2013}},
  \bibinfo{pages}{249} (\bibinfo{year}{2013}),
\eprint{1311.5010}.
%%CITATION = ARXIV:1311.5010;%%.

\bibitem[{\citenamefont{Bietenholz et~al.}(2010)}]{Bietenholz:2010}
\bibinfo{author}{\bibfnamefont{W.}~\bibnamefont{Bietenholz}}
  \bibnamefont{et~al.}, \bibinfo{journal}{Phys.~Lett.}
  \textbf{\bibinfo{volume}{B690}}, \bibinfo{pages}{436} (\bibinfo{year}{2010}),
\eprint{1003.1114}.
%%CITATION = ARXIV:1003.1114;%%.

\bibitem[{\citenamefont{Bietenholz et~al.}(2011)}]{Bietenholz:2011}
\bibinfo{author}{\bibfnamefont{W.}~\bibnamefont{Bietenholz}}
  \bibnamefont{et~al.}, \bibinfo{journal}{Phys.~Rev.}
  \textbf{\bibinfo{volume}{D84}}, \bibinfo{pages}{054509}
  (\bibinfo{year}{2011}),
\eprint{1102.5300}.
%%CITATION = ARXIV:1102.5300;%%.

\bibitem[{\citenamefont{Constantinou et~al.}(2014)}]{axialRenormalisation}
\bibinfo{author}{\bibfnamefont{M.}~\bibnamefont{Constantinou}}
  \bibnamefont{et~al.} (\bibinfo{collaboration}{QCDSF}), \bibinfo{journal}{in
  preparation}  (\bibinfo{year}{2014}).

\bibitem[{\citenamefont{Capitani et~al.}(2001)}]{Capitani:2000xi}
\bibinfo{author}{\bibfnamefont{S.}~\bibnamefont{Capitani}}
  \bibnamefont{et~al.}, \bibinfo{journal}{Nucl.~Phys.}
  \textbf{\bibinfo{volume}{B593}}, \bibinfo{pages}{183} (\bibinfo{year}{2001}),
\eprint{hep-lat/0007004}.
%%CITATION = HEP-LAT/0007004;%%.

\bibitem[{\citenamefont{Cooke et~al.}(2013)}]{Cooke:2013}
\bibinfo{author}{\bibfnamefont{A.~N.} \bibnamefont{Cooke}}
  \bibnamefont{et~al.}, \bibinfo{journal}{PoS}
  \textbf{\bibinfo{volume}{LATTICE2013}}, \bibinfo{pages}{278}
  (\bibinfo{year}{2013}),
\eprint{1311.4916}.
%%CITATION = ARXIV:1311.4916;%%.

\bibitem[{\citenamefont{Syritsyn}(2014)}]{Syritsyn:2014saa}
\bibinfo{author}{\bibfnamefont{S.}~\bibnamefont{Syritsyn}}
  (\bibinfo{year}{2014}),
\eprint{1403.4686}.
%%CITATION = ARXIV:1403.4686;%%.

\bibitem[{\citenamefont{G{\"o}ckeler et~al.}(2002)}]{Gockeler:2002}
\bibinfo{author}{\bibfnamefont{M.}~\bibnamefont{G{\"o}ckeler}}
  \bibnamefont{et~al.} (\bibinfo{collaboration}{QCDSF}),
  \bibinfo{journal}{Phys.~Lett.} \textbf{\bibinfo{volume}{B545}},
  \bibinfo{pages}{112} (\bibinfo{year}{2002}),
\eprint{hep-lat/0208017}.
%%CITATION = HEP-LAT/0208017;%%.

\bibitem[{\citenamefont{Zanotti et~al.}(2003)}]{Zanotti:2003}
\bibinfo{author}{\bibfnamefont{J.~M.} \bibnamefont{Zanotti}}
  \bibnamefont{et~al.} (\bibinfo{collaboration}{CSSM Lattice collaboration}),
  \bibinfo{journal}{Phys.~Rev.} \textbf{\bibinfo{volume}{D68}},
  \bibinfo{pages}{054506} (\bibinfo{year}{2003}),
\eprint{hep-lat/0304001}.
%%CITATION = HEP-LAT/0304001;%%.

\bibitem[{\citenamefont{Alexandrou
  et~al.}(2013{\natexlab{c}})}]{Alexandrou:2013b}
\bibinfo{author}{\bibfnamefont{C.}~\bibnamefont{Alexandrou}}
  \bibnamefont{et~al.}, \bibinfo{journal}{Phys.~Rev.}
  \textbf{\bibinfo{volume}{D87}}, \bibinfo{pages}{114513}
  (\bibinfo{year}{2013}{\natexlab{c}}),
\eprint{1304.4614}.
%%CITATION = ARXIV:1304.4614;%%.

\bibitem[{\citenamefont{Best et~al.}(1997)}]{Best:1997}
\bibinfo{author}{\bibfnamefont{C.}~\bibnamefont{Best}} \bibnamefont{et~al.},
  \bibinfo{journal}{Phys.~Rev.} \textbf{\bibinfo{volume}{D56}},
  \bibinfo{pages}{2743} (\bibinfo{year}{1997}),
\eprint{hep-lat/9703014}.
%%CITATION = HEP-LAT/9703014;%%.

\bibitem[{\citenamefont{Nakamura and St{\"u}ben}(2010)}]{Nakamura:2010}
\bibinfo{author}{\bibfnamefont{Y.}~\bibnamefont{Nakamura}} \bibnamefont{and}
  \bibinfo{author}{\bibfnamefont{H.}~\bibnamefont{St{\"u}ben}},
  \bibinfo{journal}{PoS} \textbf{\bibinfo{volume}{LATTICE2010}},
  \bibinfo{pages}{040} (\bibinfo{year}{2010}),
\eprint{1011.0199}.
%%CITATION = ARXIV:1011.0199;%%.

\bibitem[{\citenamefont{Boyle}(2009)}]{Boyle:2009}
\bibinfo{author}{\bibfnamefont{P.~A.} \bibnamefont{Boyle}},
  \bibinfo{journal}{Comput.~Phys.~Commun.} \textbf{\bibinfo{volume}{180}},
  \bibinfo{pages}{2739}
 (\bibinfo{year}{2009}).
%%CITATION = CPHCB,180,2739;%%.

\bibitem[{\citenamefont{Edwards and Joo}(2005)}]{Edwards:2004}
\bibinfo{author}{\bibfnamefont{R.~G.} \bibnamefont{Edwards}} \bibnamefont{and}
  \bibinfo{author}{\bibfnamefont{B.}~\bibnamefont{Joo}}
  (\bibinfo{collaboration}{SciDAC Collaboration, LHPC Collaboration, UKQCD
  Collaboration}), \bibinfo{journal}{Nucl.~Phys.~Proc.~Suppl.}
  \textbf{\bibinfo{volume}{140}}, \bibinfo{pages}{832} (\bibinfo{year}{2005}),
\eprint{hep-lat/0409003}.
%%CITATION = HEP-LAT/0409003;%%.

\end{thebibliography}

\end{document}